# Neutron scattering evidence for two-dimensionally coupled spin-dimerized antiferromagnetic lattice in $\alpha$-$Cu_2P_2O_7$

B. Ghanta,[1,2] K. S. Chikara,[1,2] M. Ghanathe,[3] L. Keller,[4] and D. Voneshen[5,6]

and A. K. Bera,[1,2*]

[1]*Solid State Physics Division, Bhabha Atomic Research Centre, Mumbai 400085, India;*
[2]*Homi Bhabha National Institute, Anushaktinagar, Mumbai 400094, India*
[3]*Technical University of Munich, Heinz Maier-Leibnitz Center (FRM II), 85748 Garching, Germany*
[4]*PSI Center for Neutron and Muon Sciences, CH-5232 Villigen PSI, Switzerland*
[5]*ISIS Neutron and Muon Source, Rutherford Appleton Laboratory, Didcot OX11 0QX, UK*
[6]*Department of Physics, Royal Holloway University of London, Egham TW20 0EX, UK*

[*]Corresponding author: akbera@barc.gov.in

**Abstract:**

The microscopic magnetic model of the low-dimensional quantum magnet $\alpha$-$Cu_2P_2O_7$ has remained controversial. We present a comprehensive study of its magnetic ground state and excitation spectrum using temperature-dependent inelastic neutron scattering, neutron diffraction, magnetization measurements, and comprehensive spin-wave modeling. Our results unambiguously establish $\alpha$-$Cu_2P_2O_7$ as a two-dimensionally coupled spin-dimerized antiferromagnetic (AF) lattice within the *bc* plane, with a dominant AF exchange $J_2 = 7.73 \pm 0.02$ meV (hereafter referred as 'intradimer exchange') and weaker exchange couplings $J_1$, $J_3$, and $J_4$ in the two-dimensional lattice (hereafter referred as 'interdimer exchange'), in agreement with LDA-based density functional theory and in contrast to previous GGA+U predictions. The dominant intradimer AF exchange is found between seventh-nearest-neighbor Cu–Cu ion pairs [$d_{Cu\text{-}Cu} = 5.125(3)$ Å] rather than nearest neighbor Cu–Cu ion pairs [$d_{Cu\text{-}Cu} = 3.014(1)$ Å] of the structural dimers. Weak interlayer coupling ($J_5 = 0.03 \pm 0.01$ meV) stabilizes long-range antiferromagnetic order below $T_N \approx 25$ K. We further identify a weak single-ion anisotropy, associated with the distorted $CuO_5$ polyhedra, that opens a gap in the spin-excitation spectrum and drives a field-induced metamagnetic transition. Systematic spin-wave calculations elucidate the distinct roles of interlayer coupling $J_5$ and anisotropy term $D$ in producing two distinct energy gaps at different antiferromagnetic zone centers. Complementary neutron diffraction and magnetization measurements as a function of applied magnetic field uncover a previously overlooked metamagnetic transition near 13 kOe and allow construction of the magnetic phase diagram in the $H$-$T$ plane. Our results unambiguously establish $\alpha$-$Cu_2P_2O_7$ as a realization of a two-dimensionally coupled spin-dimerized AF lattice with weak interlayer coupling and magnetic anisotropy, and thus, resolve the discrepancy in the microscopic magnetic model. Present findings also highlight the importance of extended exchange pathways and subtle anisotropic exchange interactions in governing the ground state and excitation spectrum of low-dimensional quantum magnets.

## I. INTRODUCTION

Spin-dimer systems are an important class of quantum magnets in which pairs of magnetic ions are coupled by strong antiferromagnetic exchange interactions, giving rise to a singlet ground state separated from triplet excitations by a finite spin gap [1]. These systems provide an ideal platform for exploring quantum many-body effects where the balance between intra-dimer and inter-dimer interactions drive quantum phase transitions from nonmagnetic gapped states/phases to magnetically ordered states/phases. Spin-dimer systems having relatively weak interdimer couplings allow gapped state to be tuned by external parameters such as magnetic field, pressure, or chemical substitution, leading to magnetic ordering at finite temperatures. Such perturbation-induced closing of the spin gap is often described in terms of Bose–Einstein condensation of triplon excitations [2], offering a unifying framework to study quantum criticality and collective excitations in low-dimensional magnets. Consequently, spin-dimers have attracted sustained interest as model systems for investigating quantum phase transitions, magnetic excitations, and the interplay between dimensionality, frustration, and correlations.

Materials realizing the spin-dimer model, coupled via interdimer interactions in one-dimension (1D) ($BaCu_2V_2O_8$, $Sr_{14}Cu_{24}O_{41}$) or two-dimension (2D) ($BaCuSi_2O_6$, $SrCu_2(BO_3)_2$) or three-dimension (3D) ($TlCuCl_3$, $Sr_3Cr_2O_8$), reveal diverse magnetic behaviors, including quantum entanglement ($Sr_{14}Cu_{24}O_{41}$), Bose-Einstein condensation ($TlCuCl_3$), and magnetization plateaus ($SrCu_2(BO_3)_2$) [2-8]. The natures of magnetic ground state and magnetic properties as well as their field and temperature dependences are strongly dependent on the relative strength of inter-dimer to intra-dimer couplings, spin-value, as well as anisotropy value. As the system evolves with increasing the strength of effective interdimer interactions, the energy gap reduces and leading to a long-range magnetic order at finite temperature [9,10], albeit the resulting ordered magnetic states differ from classical ones. Here, the strength, sign, as well as numbers of the competing inter-dimer exchange interactions are particularly crucial to decide the magnetic ground state; as well as temperature and magnetic field dependent magnetic properties. For example, despite having similar structural motif in $\alpha$-$Cu_2P_2O_7$ [11], $\alpha$-$Cu_2As_2O_7$ [10], and $\beta$-$Cu_2V_2O_7$ [12] (featuring same basic layered crystal structures with edge shared $CuO_x$ polyhedra separated by pyrophosphate/arsenate/vanadate groups) diverse variation in magnetic behaviors have been reported, viz., spin dimerization in $\alpha$-$Cu_2P_2O_7$, alternating spin-chains in $\alpha$-$Cu_2As_2O_7$, and a significantly different honeycomb lattice in $\beta$-$Cu_2V_2O_7$. Such significant different magnetic Hamiltonians in isostructural compounds are linked to the different natures of the interdimer couplings. Similar interdimer coupling dependent different magnetic ground states were also reported for several V-ion ($V^{4+}$: $3d^1$) based spin-1/2 dimer systems, viz., $VOSeO_3$ [13], $(VO)_2P_2O_7$ [14,15], and $CsV_2O_5$ [16], despite their crystal structural likenesses (dimers are formed by edge shared $V^{4+}O_5$ polyhedra, separated by non-magnetic $P^{5+}O_4$, $V^{5+}O_4$ and $Se^{4+}O_3$ polyhedral groups). For instance, $VOSeO_3$ has been described by weakly coupled dimer networks with small energy gap, $(VO)_2P_2O_7$ has been described by weakly coupled alternating spin-chains model, and $CsV_2O_5$ has been described by coupled alternating spin-chain model. Moreover, copper (II) selenites compounds $ACu(SeO_3)_2$ ($A$ = Hg, Cd, Ca) share a characteristic spin-dimer structure $Cu_2O_8$, formed by edge shared $CuO_5$ tetragonal pyramids, that are separated by $SeO_3$ groups. While structurally similar and having a non-magnetic singlet ground state with typical spin-dimer behavior, they exhibit different energy gaps in their magnetic spectrum, due to different strengths of inter-dimer interactions [17,18].

In similar line, despite being isostructural crystal lattice, constituted with similar quasi-2D structures, featuring Cu-$X$ ($X$=Cl, Br) sheets sandwiched between double-perovskite $LaNb_2O_7$ slabs with tetragonal ($P4/mmm$) space group, $(CuCl)LaNb_2O_7$ and $(CuBr)LaNb_2O_7$ exhibit crucially different magnetic ground states, viz., spin-singlet and long-range antiferromagnetic ordered magnetic states, respectively [19,20]. In addition, $Cu_3Zn(OH)_6Cl_2$ and $Cu_3Mg(OH)_6Cl_2$ show distinct magnetic behaviors, regardless of sharing the same crystal structure type, characterized by stacked 2D kagome layers (a network of corner-sharing triangles) of magnetic $Cu^{2+}$ ions, separated by layers containing Zn/Mg, $Cl^-$, and $OH^-$ groups. Where, $Cu_3Zn(OH)_6Cl_2$ remains a quantum spin liquid down to very low temperatures [21,22], while $Cu_3Mg(OH)_6Cl_2$ exhibits weak ferromagnetic ordering [23]. Hence, in general, the nature of constituted magnetic exchange interactions that defines the effective Hamiltonian of the magnetic systems play a crucial role in the magnetic properties of the low-dimensional quantum spin systems. Especially, the magnetic exchange interactions in insulators, constituted by superexchange interactions, are accepted to be defined by Goodenough-Kanamori-Anderson (GKA) rules, i.e., by virtue of the values of bond-angles and bond-lengths involved in the exchange pathways [24-27]. However, these rules are not sufficient, especially for long-range couplings that involve more than one ligand atom, leading to incorrect definition of magnetic exchange interactions, subsequently, the magnetic Hamiltonian and the dimensionality of the magnetic system [11].

Our present interest is on the coupled spin-1/2 dimer system $\alpha$-$Cu_2P_2O_7$ for which the spin Hamiltonian remains controversial. As per the report by Janson *et al.*, [11] the magnetic model of $\alpha$-$Cu_2P_2O_7$ consists of two-dimensionally coupled spin-1/2 dimers (within the *bc* plane), where sizable interdimer couplings enforce the antiferromagnetic long-range ordering. However, the magnetic dimers are proposed to form between the $7^{th}$ nearest neighbour (NN) Cu-spins, separated by a significantly large distance of $d_{Cu-Cu}$ =5.125(3) Å, as compared to the structural dimer formed by the $1^{st}$ NN Cu-spins [$d_{Cu-Cu}$ =3.014(1) Å]. The strongest exchange interaction is proposed to be occurred through Cu-O-O-Cu super-super exchange pathways, and inconsistent with the concepts of GKA rules. In sharp contrast, a recent work by Yang *et al*. [28] on $\alpha$-$Cu_2P_2O_7$, reported a quasi-1D zig-zag spin-chain model (running perpendicular to the *bc*-plane) consists of stronger exchange interaction between the NN Cu-ions (structural dimers), following the GKA rules. Such controversy demands an experimental verification of the magnetic lattice of $\alpha$-$Cu_2P_2O_7$, especially, by INS measurement which provides momentum-resolved excitation spectra for a wide range of energies, suitable for precise determination of exchange interactions. Moreover, the electronic band structure calculations employing local density approximation (LDA) by Janson *et al*. [11] predicted a metallic state in contrast to the semiconducting state with band gap value of ~ 3 eV predicted by Yang *et al*. [28] using the generalized gradient approximation (GGA) combined with the Hubbard-U correction (GGA+U) calculations, which needs a reverification.

In the present study, by performing in-depth temperature dependent INS measurements and by carried out a comprehensive spin-wave analyses, we unambiguously clarify the magnetic Hamiltonian of $\alpha$-$Cu_2P_2O_7$ is as a two-dimensionally coupled magnetic dimer model (within the *bc* plane) in agreement with the LDA based DFT predictions by Janson *et al*. [11] and in contrast to the GGA+U based DFT calculations by Yang *et al*., [28]. The 2D coupled magnetic dimer model is determined with intradimer exchange interaction $J_2$ =7.73 ± 0.02 meV, and in-plane interdimer exchange interactions $J_1$ = 2.8 ± 0.03 meV, $J_3$ = 3.61 ± 0.01 meV and $J_4$ = 3.23 ± 0.05 meV. Here, the magnetic dimers (where $J_2$ =7.73 meV) are formed with the $7^{th}$ NN Cu-Cu ions separated by a longer distance [$d_{Cu-Cu}$ = 5.125(3) Å], rather than between the NN Cu-Cu ions [$d_{Cu-Cu}$ = 3.014(1) Å] (where $J_1$ = 2.8 meV). The 2D magnetic layers are weakly coupled ($J_5$ = 0.03 ± 0.01 meV) to form a 3D magnetic lattice and resulting into the observed magnetic transition to ordered 3D AFM ground state below ~ 25 K. Our results also reveal the presence of a weak but finite single-ion-anisotropy ($D$) for the $Cu^{2+}$ spins, residing within the distorted $CuO_5$ polyhedral environment, resulting into energy gaps in the spin excitations spectra as well as a field induced metamagnetic transition under ~ 13 kOe. Our systematic SW calculations shed light on the roles of weak inter-planar coupling $J_5$ and single-ion-anisotropy $D$ on the observed two different energy gaps at different AFM zone centres. We have also revisited the magnetic structure by a comprehensive temperature dependent powder neutron diffraction study and provide explanation for the preservation of the unit cell centrosymmetry under the AFM ordering. Detailed temperature and field dependent dc-magnetization studies along with field dependent neutron powder diffraction study have been carried out to shed light on the previously ignored field induced metamagnetic transition as well as to determine the magnetic phase diagram in the $H$-$T$ plane. Our study, thus, illustrate a realization of a two-dimensional spin-dimer system in $\alpha$-$Cu_2P_2O_7$ and resolve the discrepancy in the microscopic magnetic model.

## II. EXPERIMENTAL and COMPUTATIONAL DETAILS

Polycrystalline samples of copper pyrophosphate $\alpha$-$Cu_2P_2O_7$ were synthesized by the solid-state reaction method. Stoichiometric mixture of high purity (Sigma-Aldrich, > 99.99%) precursors (CuO and $(NH_4)_2HPO_4$) was well ground using agate mortar and pestle. Then, homogeneously mixed powder was sintered at 1073 K in air for total 72 hours with several intermediate grindings. The phase purity of $\alpha$-$Cu_2P_2O_7$ was confirmed by x-ray diffraction (with Cu $K_\alpha$ radiation).

The room temperature crystal structure of $\alpha$-$Cu_2P_2O_7$ was studied by neutron powder diffraction using the PD-1 diffractometer ($\lambda$= 1.094 Å) at Dhruva Research Reactor, BARC, Mumbai, India [29]. The measured diffraction patterns were analysed by Rietveld refinement method with the help of FULLPROF computer program [30]. Low temperature (2−50 K) neutron powder diffraction measurements were carried out using the high-intensity cold-neutron powder diffractometer DMC ($\lambda$ = 2.4586 Å) at the Paul Scherrer Institute (PSI), Switzerland [31]. Additional magnetic field dependent neutron powder diffraction measurements were carried out over 0−60 kOe at base temperature (2 K) using the MA6 cryo-magnet. For DMC measurements, a sintered (at 1073 K) cylindrical-shaped rod of powder samples of $\alpha$-$Cu_2P_2O_7$ was used to minimize the particle realignment in the presence of external magnetic field. Magnetic symmetry analyses, to determine the magnetic structure, were carried out using the MAXMAGN program available in the Bilbao Crystallographic Server [32,33]. Magnetic refinement was carried out using the FULLPROF computer program.

The dc-magnetization and ac-susceptibility measurements were carried out using the M/S Cryogenic Co. Ltd., UK, make multiproperty measurement system. Temperature-dependent dc-susceptibility ($\chi_{dc}$) curves were measured under magnetic fields of 1, 2, 3, 4, 5, 6, 7, 8, 9, 10, 11, 12, 15 and 20 kOe, and temperature-dependent ac-susceptibility ($\chi_{ac}$) curve was measured at ac magnetic field of 1mT (988 Hz) over 2–300 K. Field-dependent isothermal magnetization $M(H)$ curves were measured at 2, 4, 7.5, 10, 12.5, 14, 16, 18, 20, 22 and 30 K over 0−40 kOe.

The inelastic neutron scattering (INS) measurements were carried out on powder samples (~ 20 gm) of $\alpha$-$Cu_2P_2O_7$ using the high-count rate time-of-flight neutron spectrometer MERLIN at the ISIS neutron and muon source, Rutherford Appleton Laboratory, Didcot, United Kingdom [34]. The large detector bank coverage of the MERLIN spectrometer in both the horizontal (~180°) and vertical (±30°) scattering planes allows measurement over large-$Q$ regions of $S(Q,\omega)$ space. For the INS measurements, the powder sample was packed in a thin aluminum (Al) foil, rolled into annular cylindrical form, and inserted inside a thin-walled cylindrical Al can. The sample filled aluminum can was then mounted into the instrument and cooled in He-4 exchange gas using a closed-cycle refrigerator for low-temperature measurements. The straight Gd slit package was used in the Fermi chopper, which was phased (at a rotation speed of 400 Hz) to allow the recording of spectra with incident neutron energies ($E_i$) of 11.3, 16.9, 28.2, and 55.9 meV, simultaneously, via the rep-rate multiplication method. The INS spectra were recorded for ~2 h at each temperature of 7, 20, 50 and 100 K. The INS data were plotted and analyzed using the MSLICE program implemented in the MANTID software package [35]. The SPINW computer program was used for the spin-wave simulations [36].

The first-principles density-functional theory (DFT) calculations were performed using the VASP package [37]. The calculations employed the full-potential projector augmented-wave (PAW) method and the generalized gradient approximation (GGA) as implemented in VASP [38-40]. Correlation effects in the GGA+U scheme were included following the method proposed by Liechtenstein *et al*. [41]. The crystal structure was optimized using a 1×1×1 unit cell containing 44 atoms. Brillouin-zone integrations were carried out with a 4×4×4 Monkhorst-Pack *k*-point mesh and a plane-wave cut-off energy of 500 eV. Local electron–electron interactions were treated within the GGA+U framework to account for the strong on-site Coulomb repulsion of localized *d* electrons, which are inadequately described by standard LDA or GGA. We adopted $U$ = 9 eV and $J$ = 1 eV for Cu-ions, and $U$ = 4 eV and $J$ = 0.5 eV for O and P -ions, adopted from the earlier report for similar Cu-based compounds [42]. The experimental lattice parameters were used as the starting point and subsequently relaxed to obtain the optimized structure.

## III. RESULTS AND DISCUSSION

### A. Crystal structural properties

The crystal structure of $\alpha$-$Cu_2P_2O_7$ has been investigated by neutron diffraction at room temperature. The Rietveld analysis of room temperature neutron diffraction pattern (with the nuclear structure having *C2/c* symmetry) is shown in Fig. 1(a). The Rietveld analysis reveals the $\alpha$-$Cu_2P_2O_7$ crystallizes in the $\alpha$-phase (space group: *C2/c*, No. 15) belonging to the thortveitite structure family with the general formula $M_2P_2O_7$, where $M$ = divalent transition metal cations ($M$ = Mg, Mn, Cu, Ni, Zn, and Co).

The refined lattice parameters, fractional atomic coordinates and site occupancy of atoms, obtained from the analysis of neutron diffraction pattern are given in Table I. The refined values of lattice parameters at room temperature are $a$ = 6.883(1) Å, $b$ = 8.115(2) Å, $c$ = 9.166(3) Å, and $\beta$ = 109.56 (2) °, which are in good agreement with the previously reported values in ref. [43]. The crystal structure of $\alpha$- $Cu_2P_2O_7$ is composed of $Cu_2O_8$ structural dimers (formed by two edge-sharing $CuO_5$ square pyramids) which are separated by nonmagnetic $PO_4$ tetrahedra. The magnetic $Cu^{2+}$ ($3d^9$, $S$=1/2) transition metal-ion is coordinated with five O atoms forming a distorted $CuO_5$ square pyramidal geometry [Fig. 1(d)]. The apical Cu–O3 bond length is significantly longer [2.339 Å; ~ 20% higher] than that of the average bond length of the four bond lengths (in the range of 1.88–1.9 Å) in the basal plane of the distorted $CuO_5$ square pyramid [Fig. 1(d) and Table II], attributed to the Jahn–Teller distortion [44]. The non-magnetic $P^{5+}$ ($2p^6$, $S$ = 0) ion is coordinated with four O atoms forming a distorted $PO_4$ tetrahedron [Fig. 1(e) and Table II].

The edge-shared two $CuO_5$ pyramids form $Cu_2O_8$ structural dimers between two nearest neighbour (NN) Cu-Cu ion pair [$d_{\text{Cu-Cu}}$ = 3.014(1) Å and Cu–O2–Cu angle: 99.9(4) °], as shown in Fig. 1(c). A schematic representation of the possible exchange interactions through the superexchange Cu–O–Cu and super-superexchange Cu–O–O–Cu pathways within the *bc* plane is shown in Fig. 1(b). Within the *bc* plane, structural dimers are coupled through $PO_4$ tetrahedra [Fig. 1(b)] to form a 2D lattice of coupled spin-dimers, where the interdimer exchange couplings are between the $7^{th}$-$9^{th}$ NN Cu-Cu ion pairs. As described in the Introduction section, a controversy present whether the strongest exchange interaction is $J_1$ (between the Cu-pairs within the structural dimer; i.e., between $1^{st}$ NN Cu-pairs) or $J_2$ (between the Cu-pairs separated by a larger distance of $d_{\text{Cu-Cu}}$ = 5.125(3) Å; i.e., between $7^{th}$ NN Cu-pairs). We resolve this discrepancy by an experimental INS study (presented later in Sec. III F). On the other hand, perpendicular to the layers (along the *a*-axis), the couplings between the dimers (between the $2^{nd}$-$6^{th}$ NN Cu-Cu ion pairs) occur via $P_2O_7$ groups and/or via a Cu–O3–Cu pathway. Such interplanar couplings are predicted to be significantly weaker than that of the in-plane exchange couplings [11].

### B. Macroscopic magnetic properties

The temperature dependent $\chi_{dc}$ curve measure under a magnetic field of 1 kOe exhibit a broad maximum at ~ 55 K and a weak kink at around $T_N \approx 22$ K [Fig. 2] as reported before in ref. [11]. The weak kink at around $T_N \approx 22$ K becomes prominent at higher applied magnetic field above ~ 2 kOe [Fig. 2(b)]. The temperature dependent $\chi_{ac}(T)$ curve also yields a broad maximum at ~ 55 K [Inset of Fig. 2(a)]. Such broad maximum indicates the signature of an onset of short-range spin-spin correlations in $\alpha$-$Cu_2P_2O_7$. In the high temperature region (> 150 K), the $1/\chi_{dc}(T)$ curve at 1 kOe follows the Curie-Weiss (CW) law $\chi_{dc}(T) = C/(T-\theta_{CW})$, where C is the Curie-Weiss constant and $\theta_{CW}$ is the Curie-Weiss temperature, owing to paramagnetic regime [Fig. 2(a)]. The linear fit to the inverse susceptibility curve yields an effective paramagnetic moment of 1.91 $\mu_B$, which is in good agreement with that reported in ref. [11]. The obtained effective paramagnetic moment is slightly larger than the spin-only effective magnetic moment i.e., $\mu_{eff} = g\sqrt{[S(S+1)]}\,\mu_B = 1.73\,\mu_B$ for spin $S$ = 1/2 of $Cu^{2+}$ ion. The enhancement beyond the spin-only contribution can be attributed to larger value of the *g*-factor ($g_{powder}$ = 2.2; $g_{\perp}$ ~ 2.1 and $g_{\parallel}$ ~ 2.4 [11]) than 2.0, arising from crystal field anisotropy or unquenched orbital contributions [45]. The measurement of $\chi_{dc}(T)$ curve under high field (≥ 2 kOe) shows a sharp peak at ~ 22 K [Fig. 2(b)], confirms the presence of long-range magnetic ground state, and the peak temperature decreases with the increasing magnetic field. The peak in the $\chi_{dc}(T)$ curve ~22 K turns into a steep upturn beyond a critical magnetic field $H_C$ ~13 kOe, indicating a field-induced magnetic phase transition from AFM ground state to a field induced intermediate state. The temperature and magnetic field effects on the AFM ordering is further investigated by Fisher's specific heat [$d(\chi_{dc}T)/dT$] which generally reveals an anomaly at the magnetic ordering temperature. The plots of Fisher's specific heat [Fig. 2(c)] show an anomaly at ~ 22 K for magnetic field 2 kOe and above, suggesting the onset of long-range magnetic ordering. It is evident that the anomaly progressively smears out with increasing the magnetic field.

To investigate the effect of magnetic field on the AFM ordered state of $\alpha$-$Cu_2P_2O_7$ further, the field-dependent isothermal magnetization *M* (*H*) curves were measured at selected temperatures over 2-30 K [Fig. 2(d)]. At 30 K (above the $T_N \approx$ 22 K), the *M* (*H*) curve shows a linear behaviour. Below the $T_N$, the *M* (*H*) curve shows almost linear behaviour with a slope change ~13 kOe, confirming a field-induced metamagnetic transition, in agreement with that reported in ref. [46]. The metamagnetic transition at $H_C$ ~13 kOe is clearly demonstrated by an anomaly in the *dM*/*dH* curve [Fig. 2(f)]. The *M*(*H*) curves do not show any hysteresis or remanent magnetization down to lowest measured temperature of 2 K. The observed value of magnetization at 2 K is *M* ~ 0.037 $\mu_B/Cu^{2+}$ under 40 kOe [2(e)], which is about 3.7% of the expected saturation magnetization (~1 $\mu_B/Cu^{2+}$), suggesting the saturation field is high for the present compound. In fact, the saturation field is calculated to be ~ 1700 kOe by QMC method for this compound [11]. With increasing temperature, the anomaly in the *dM*/*dH* broadens progressively and disappears above the $T_N$, suggesting the field-induced metamagnetic transition is related to the magnetically ordered state. A magnetic phase diagram in the *H*-*T* plane [Fig. 2(g)] has been constructed from the anomalies in the *M*(*H*) and $\chi_{dc}(T)$ curves, revealing three distinct regions; (I): an AFM ordered state, (II): a field-induced phase, and (III): a short-range ordered (SRO) phase above the $T_N \approx$ 22 K. The detailed microscopic nature of these phases is investigated by neutron diffraction and the results are discussed in the next sections.

### C. Magnetic ground state and magnetic symmetry analysis

The microscopic nature of spin-spin correlations and ground state magnetic structure have been investigated by temperature-dependent neutron diffraction study [Fig. 3(a)]. The diffraction patterns look similar over 50-2 K except an appearance of weak magnetic Bragg peaks below 25 K. This confirms that the symmetry of the nuclear structure remains same (Monoclinic space group *C2*/*c*) down to lowest measured temperature of 2 K. The measured neutron diffraction pattern at 30 K (above the $T_N$) along with simulated pattern considering the nuclear phase alone with the space group *C2*/*c* is shown in Fig. 3(c). The pure magnetic diffraction patterns were estimated by subtracting the nuclear background (measured diffraction pattern at 30 K) [Fig. 3(b)]. The appearance of additional magnetic Bragg peaks in neutron diffraction patterns ($Q$ = 1.26, 1.59, 1.71, 2.20, 2.38, and 2.84 Å$^{-1}$) below the $T_N \approx 25$ K [Figs. 3(a) and 3(b)] reveals a long-range AFM order state as concluded from the magnetization data. It is noticeable that intensities of magnetic Bragg peaks are very weak as compared to the nuclear Bragg peaks, even at 2 K.

All the magnetic Bragg peaks can be indexed with a single commensurate magnetic propagation vector $\boldsymbol{k}$ = (0,0,0). Magnetic symmetry analyses were performed using the program MAXMAGN, available in the Bilbao Crystallographic server [32,33]. The symmetry analyses reveal four possible maximal magnetic space groups (MSGs): *C2*/*c* (No. 15.85), *C2*'/*c* (No. 15.87), *C2*/*c*' (No. 15.88), and *C2*'/*c*' (No. 15.89) for the parent space group *C2*/*c* (No. 15) and correspond to the propagation vector $\boldsymbol{k}$ = (0,0,0). The maximal magnetic space group *C2*/*c* (No. 15.85), contains purely spatial symmetries without having time-reversal symmetry. In contrast, all other three space groups contain time reversal symmetry. For *C2*/*c*', *C2*'/*c*, and *C2*'/*c*', the time-reversal symmetry is associated with the *c*-glide operation, the 2-fold rotation, and both the 2-fold rotation and the *c*-glide operation, respectively.

Wyckoff positions and corresponding components of magnetic moment for the magnetic structure corresponding to four space groups are given in Table III. Within the unit cell, there are 8 Cu positions (8*f* site) which are related by spatial symmetry operations. The symmetry-related positions (-*x*, *y*, -*z* +1/2), (-*x*, -*y*, -*z*) and (*x*, -*y*, *z* +1/2) are generated from the reference position (*x*, *y*, *z*) by a 2-fold rotation about *b*-axis with a 1/2-translation along *c*-axis {$2_{010}$ | 0 0 1/2}, inversion {-1 | 0}, and *c*-glide ⊥ *b*-axis {$m_{010}$ | 0 0 1/2}, respectively. The inversion {-1 | 0} symmetry comes from the combined operations of *c*-glide ⊥ *b*-axis {$m_{010}$ | 0 0 1/2} and 2-fold rotation about *b*-axis with a ½-translation along *c*-axis {$2_{010}$ | 0 0 1/2}. Remaining four positions in the unit cell are generated by *C*-centering translation {1/2 1/2 0} to each of the above four positions [Table III]. It is evident from Table III that all three moment components ($m_x$, $m_y$, $m_z$) are symmetrically allowed for all the 8 Cu positions (under 8*f* site).

To determine the magnetic structure, the obtained magnetic crystallographic information file (mCIF) from MAXMAGN was used for the analyses of the measured neutron diffraction pattern using the FULLPROF program. Four individual magnetic models were constructed based on the four symmetry allowed magnetic space groups. Independent Rietveld refinement analyses were performed for the four magnetic models. The calculated patterns for all four models are compared with the pure magnetic diffraction pattern (i.e., 2 K-30 K pattern) [Figs. 3(f-i)]. We found that the best agreement was achieved for the magnetic space group *C2/c'* [Fig. 3(i)]. For a comparison, we have also shown the as-measured [Fig. 3(d)] and pure magnetic [Fig. 3(e)] diffraction patterns along with the calculated diffraction patterns considering the nuclear (*C2/c*) + magnetic phases (*C2/c'*) and only magnetic phase (*C2/c'*), respectively. The corresponding magnetic structure is a commensurate AFM structure as shown in Fig. 4(a). For the space group *C2/c'*, the direction of magnetic moment components at positions ($x$, $y$, $z$), ($-x$, $y$, $-z$ +1/2), ($x$, $y$, $z$)+{1/2 1/2 0}, and ($-x$, $y$, $-z$ +1/2)+{1/2 1/2 0} are same due to the absence of time reversal symmetry. In contrast, direction of magnetic moment components at positions ($-x$, $-y$, $-z$), ($x$, $-y$, $z$ +1/2), ($-x$, $-y$, $-z$)+{1/2 1/2 0}, and ($x$, $-y$, $z$ +1/2)+{1/2 1/2 0} reverse due to the presence of time-reversal symmetry associated with the *c*-glide symmetry operation [Table III]. The projection of the magnetic structure in the *bc* plane is shown in Fig. 4(c) which yields an antiferromagnetic spin arrangement for both the structural dimers [$d_{\text{Cu-Cu}}$ = 3.014(1) Å] as well as the proposed magnetic dimer [$d_{\text{Cu-Cu}}$ = 5.125(3) Å] [Fig. 4(c)]. Each of the spins of a given dimer (either structural or magnetic dimer) is coupled antiparallel to the spins from the given dimer and neighbouring dimers [Fig. 4(c)].

The Rietveld analyses of the magnetic diffraction pattern, measured at 2 K, results a site ordered magnetic moment of $m_{\text{Cu}}$ = 0.670(5) $\mu_{\text{B}}$/$Cu^{2+}$, which is significantly reduced (33%) from the theoretically expected value of 1$\mu_{\text{B}}$ for the $Cu^{2+}$(3$d^9$, $S$ = 1/2) ion, i.e., g$\mu_{\text{B}}S$ = 1$\mu_{\text{B}}$. The determined magnetic moment components along $a$, $b$, and $c$ axes are: $m_a$ = 0.448(4) $\mu_{\text{B}}$, $m_b$ = 0.01(1) $\mu_{\text{B}}$, and $m_c$ = 0.674(3) $\mu_{\text{B}}$ at 2 K. Therefore, the magnetic moment lies entirely within the *ac* plane [Fig. 4(c)] having a tilt from the *c*-axis by 39(5)° [Fig. 4(b)], suggesting an easy-axis approximately along the [201] direction [47]. The temperature dependence of ordered magnetic moment ($m_{\text{Cu}}$) of $Cu^{2+}$ ions, as derived from the refinement of neutron diffraction patterns at individual temperatures, is shown in Fig. 4(d). With increasing temperature, the $m_{\text{Cu}}$ progressively reduces and becomes zero at the $T_{\text{N}}$. For a second-order magnetic phase transition, like the present case, the order parameter is a continuous function of temperature, and follows a power law around the phase transition. The experimentally determined $m_{\text{Cu}}$ vs temperature curve is fitted with a power law function [48]

$$m_{\text{Cu}}(T) = A(T_{\text{N}} - T)^{\beta} \quad (2)$$

where $A$ is a proportionality constant, $\beta$ is the critical exponent. The fitted curve over the temperature range $0.4 < T/T_{\text{N}} < 1$ is shown by the solid red line in Fig. 4(d). The fitting yields $\beta$ = 0.369(3) and $T_{\text{N}}$ = 24.89(2) K. The theoretically expected values of the critical exponent $\beta$ are 0.125 for 2D Ising, 0.326 for 3D Ising, 0.367 for the 3D Heisenberg model [49], and 0.5 for mean-field theory [45], respectively. The fitted value of $\beta$ is close to the value for 3D Heisenberg model revealing an almost isotropic nature of the spins in $\alpha$-$Cu_2P_2O_7$. In agreement to our results, the electron spin resonance (ESR) study of $\alpha$-$Cu_2P_2O_7$ reported a powder-averaged $g$ value of 2.2 ($g_{\perp}$ ~ 2.1 and $g_{\parallel}$ ~ 2.4) [11], indicating a weak g-factor anisotropy. Although there is a presence of a weak $g$ anisotropy ($g_{\perp} \neq g_{\parallel}$) in $\alpha$-$Cu_2P_2O_7$, however, not sufficient to change the universality class from the isotropic 3D Heisenberg model.

### D. Theoretical insights of magnetic ground state and electronic properties

To further elucidate the magnetic ground state and electronic properties, we employed DFT calculations. The lattice parameters obtained from relaxation with GGA approximation are listed in Table I, and found to be in good agreement with the experimentally determined values. These parameters were used for all subsequent calculations. The stability of the magnetic ground state was investigated through the total-energy calculations for spin configurations associated with each of the four magnetic space groups derived from symmetry analysis with $\boldsymbol{k}$ = (0,0,0) corresponding to the primitive 1×1×1 cell i.e., *C2/c* (No. 15.85), *C2'/c* (No. 15.87), *C2/c'* (No. 15.88), and *C2'/c'*(No. 15.89) [Table IV]. Initial GGA calculations identify the collinear spin configurations corresponding to the magnetic space group *C2/c'* as the lowest-energy state. The GGA calculations result a local Cu magnetic moment value of 0.605 $\mu_B$ (with the default moment direction along the *c*-axis). Subsequent GGA+U calculations also confirm the lowest energy for the magnetic ground state with the magnetic space group *C2/c'*. The magnetic ground state with the magnetic space group *C2/c'* determined from the present DFT calculations is in agreement with that determined from our experimental NPD data [Section IIIC] as well as earlier report from DFT calculations [11]. The local Cu magnetic moment value is estimated by the GGA+U calculations to be 0.83 $\mu_B$ (with the default moment direction along the *c*-axis) in agreement with that reported earlier [28]. Additional GGA+U calculations, by allowing $Cu^{2+}$ magnetic moment components along all the three crystallographic directions, yield a total moment value of 0.71 $\mu_B$/$Cu^{2+}$ with components $m_a$ = 0.471 $\mu_B$, $m_b$ = 0.001 $\mu_B$, and $m_c$ = 0.709 $\mu_B$ which are in close agreement with the experimentally determined values of $m_a$ = 0.448(4) $\mu_B$, $m_b$ = 0.01(1) $\mu_B$, and $m_c$ = 0.674(3) $\mu_B$, respectively. A small induced magnetic moment (approximately 0.02 $\mu_B$) on oxygen ions was also found, attributed to Cu–O hybridization.

The total and projected density of states (DOS and PDOS, respectively) obtained from the GGA+U calculations are shown in Fig. 5. The valence band maximum is dominated by O–*p* states with significant Cu–*d* hybridization, while the conduction band primarily consists of Cu–*d* states. The calculated band structure reveals an insulating behavior with a band gap of ~3.1 eV, consistent with previous reports by Yang *et al*. [28] using the same GGA+U DFT method as well as experimentally reported semiconducting behavior [50].

### E. Magnetic field dependent spin-spin correlation

To shed light on the microscopic nature of the magnetic field-dependent metamagnetic transition at $H_C$ ~13 kOe, magnetic field-dependent neutron diffraction measurements were carried out at 2 K on a pressed rod of $\alpha$-$Cu_2P_2O_7$ powders. The measured neutron diffraction patterns under magnetic field over 0-60 kOe are shown in Fig. 6(a). The comparison of the neutron diffraction patterns measured in zero-field and under 60 kOe (highest applied magnetic field) is shown in Fig. 6(b). No grain reorientation is evident from the constant intensities of the nuclear Bragg peaks under the magnetic field. This is particularly helpful for the analyses of the magnetic diffraction patterns under applied magnetic field. The magnetic field-dependence of the pure magnetic intensity of the strongest magnetic Bragg peak (-1,1,1)+***k*** (pure magnetic patterns are estimated by subtracting the measured pattern at 30 K under zero applied field, considering as the nuclear background) is shown in Fig. 6(c) which shows a sudden decrease in magnetic Bragg peak intensity at $H_C$ ~13 kOe, reflecting the field-induced metamagnetic transition, consistent with the isothermal magnetization $M(H)$ curve [Fig. 2(e)]. For further details, the square root of the integrated intensity of the magnetic Bragg peak (-1,1,1)+***k*** [proportional to the magnetic moment] is shown as a function of applied magnetic field in Fig. 6(d), which shows a sudden decrease in the value by ~ 10 % at $H_C$ = 13 kOe and then remains unchanged up to the maximum applied magnetic field of 60 kOe. On the other hand, the intensity of the nuclear Bragg peaks remains unchanged under applied magnetic field [insets of Fig. 6(d)]. The above observations suggest a spin-reorientation transition within the AFM ordered state possibly driven by local single ion anisotropy. Presence of a weak single ion anisotropy is confirmed from our present INS study (discussed below) as well as reported ESR measurements [11].

### F. Spin-wave excitations and Spin-Hamiltonian

To determine the spin-Hamiltonian of the studied compound $\alpha$-$Cu_2P_2O_7$ experimentally and resolve the conflicting theoretical predictions for spin-models, we have performed inelastic neutron scattering (INS) measurements using the time-of-flight neutron spectrometer MERLIN. As pointed out in the introduction, two spin-models were proposed for the studied compounds; (i) a 2D coupled magnetic dimer model (within the *bc* plane), [11], having strongest dimer interaction between $7^{th}$ NN Cu-spins [$d_{Cu\text{-}Cu}$ = 5.125(3) Å] along the *b*-axis, and (ii) a quasi-1D zig-zag spin-chain model (running perpendicular to the *bc*-plane) consists of stronger exchange interaction between the NN Cu-ions (structural dimers) [$d_{Cu\text{-}Cu}$ = 3.014(1) Å] [28]. Here, we aim to distinguish these two models experimentally by INS study and a comprehensive spinwave analyses. The two-dimensional color-coded INS intensity maps, measured at 7, 20, 50 and 100 K with different $E_i$ =11.3, 16.9, 28.2, and 55.9 meV, are shown in Fig. 7 as a function of momentum transfer $|Q|$ and energy transfer ($E$). The sharp reduction of the intensity of the excitations above the $T_N$ = 25 K along with the disappearance of the energy (excitation) gap confirms the spin waves excitations originating from the long-range AFM ordered state of $\alpha$-$Cu_2P_2O_7$. The persistence of weak magnetic excitations at 50 K, well above the $T_N$, indicates the presence of short-range spin-spin correlations in $\alpha$-$Cu_2P_2O_7$, consistent with the magnetic susceptibility results [Fig. 2(a)]. Presence of short-range spin-spin correlations much above the long-range ordering temperature ($T_N$ = 25 K) are hallmark of low-dimensional magnetic systems and have been reported in several related compounds [51-53]. Such short-range spin-spin correlations are predominantly present within the $bc$-plane of $\alpha$-$Cu_2P_2O_7$ where the strong intraplanar exchange couplings are found (Table V and discussed in details later), revealing the low-dimensional magnetic behaviour of the studied compound $\alpha$-$Cu_2P_2O_7$.

The major features of the spin-wave excitation spectra at 7 K (the lowest measured temperature) are (i) dispersive nature of magnetic excitations, with a strong intensity band centres around ~10–13 meV, (ii) absence of magnetic excitations above 13 meV, (iii) the minima of the dispersions at ~ $|Q|$ =0.78 and $|Q|$ =1.62 $Å^{-1}$ corresponding to the reciprocal lattice points (0,1,0) and (1,1,1), respectively [Fig. 7(i)], and (iv) the presence of energy gaps with different values of ~1.17 meV and 0.80 meV at reciprocal lattice points (0,1,0) and (1,1,1), respectively [Fig. 7(i)]. To quantitatively interpret the experimentally observed magnetic spectra of $\alpha$-$Cu_2P_2O_7$, we have simulated spin-excitation spectra using the SPINW program and compared with the experimental powder INS spectra. Even though directional information is lost in powder INS spectra, the preserved energy dependent singularities in the density of states provide characteristic of the Spin-Hamiltonian(sign and strength of the exchange interactions), thus distinguish the proposed two spin-models. We have considered total five different exchange interactions i.e., $J_1$, $J_2$, $J_3$, $J_4$, and $J_5$ [Figs. 1(b) and 4(c)] and the spin Hamiltonian is expressed as:

$$H = \sum_{k=1}^{N/2}[J_1(\boldsymbol{S}_{k,1}\cdot\boldsymbol{S}_{k,2})+J_2(\boldsymbol{S}_{k,1}\cdot\boldsymbol{S}_{k\text{-}1,2})]+\sum_{k,l}[\,J_3(\boldsymbol{S}_{k,1}\cdot\boldsymbol{S}_{l\text{-}1,1})+J_4(\boldsymbol{S}_{k,2}\cdot\boldsymbol{S}_{l,2})]$$

$$+\sum_{k,m}J_5(\boldsymbol{S}_{k,1}\cdot\boldsymbol{S}_{m,2})+\sum_{k,i=1,2}^{N/2}D(S_{k,i}^{z})^2 \qquad (3)$$

where $\boldsymbol{S}_{k,i}$ is quantum spin operator at $i^{th}$ (= 1 and 2) site of the $k^{th}$ structural dimer [for shortest Cu-Cu distance of 3.014(1) Å]. The term $\boldsymbol{S}_{k,i}\,.\,\boldsymbol{S}_{l,j}$ represents interaction between the $i^{th}$ site of the $k^{th}$ structural dimer and $j^{th}$ site of the $l^{th}$ structural dimer. The parameters $J_1$ is the NN exchange interaction [$d_{Cu\text{-}Cu}$ = 3.014(1) Å], and $J_2$–$J_4$ are the in-plane exchange interactions (between 7th, 8th and 9th NN Cu-ion pairs), $J_5$ is the inter-planar exchange interaction (between the 2nd NN Cu-Cu ion pairs) and $D$ represents a single-ion anisotropy. The exchange interactions $J_1$ and $J_2$ are along the $b$-axis, while $J_3$ and $J_4$ exchange interactions are along the diagonal directions of the $bc$ plane.

The spin wave simulation is based on the ground state AFM spin structure as determined from the low-temperature neutron diffraction study [Fig. 3], having the spin-moments within the *ac* plane. The spin-wave calculations assumed a magnetic form factor corresponding to $Cu^{2+}$ ions and a spin value of $S = 1/2$. The solution of the Hamiltonian was tested over a wide range of parameter spaces ($J_1$, $J_2$, $J_3$, $J_4$, $J_5$, and $D$) compatible with the observed magnetic structure. By a systematic adjustment of values of all the exchange parameters and anisotropy, a good solution is obtained with the parameters as listed in Table V; i.e., $J_1 = 2.8 \pm 0.03$ meV (AFM), $J_2 = 7.73 \pm 0.02$ meV (AFM), $J_3 = 3.61 \pm 0.01$ meV (AFM), $J_4 = 3.23 \pm 0.05$ meV (AFM), $J_5 = 0.03 \pm 0.01$ meV (AFM), and $D = -0.07 \pm 0.02$ meV. The corresponding simulated powder-averaged excitation spectrum (convoluted with experimental resolution function of MERLIN with $E_i = 28.2$ meV) is presented in Fig. 8(b) and compared with the experimental spectrum measured at 7 K with $E_i = 28.2$ meV [Fig. 8(a)]. The simulated spectrum nicely reproduces all the key features of the experimental INS spectrum. Quantitative agreements are shown through constant-energy (integrated over $\Delta E = 2$–$4$ meV) and constant-$|Q|$ (integrated over $|Q| = 0$–$3$ Å$^{-1}$) cut plots [Figs. 8(c) and 8(d)].

The agreements for energy gaps are further demonstrated through experimental low-energy and high-resolution INS spectrum (measured with $E_i = 11.3$ meV at 7 K) and calculated spectrum, convoluted with corresponding experimental resolution function for $E_i = 11.3$ meV [Figs. 8(e-h)]. The experimental spectrum [Fig. 8(e)] reveals two different energy gap values of ~1.17 meV and 0.80 meV for the AFM zone centres at the reciprocal points (0,1,0) and (1,1,1) [ $|Q| = 0.78$ and $|Q| = 1.62$ Å$^{-1}$], respectively, which are in excellent agreement with the simulated spectrum based on the determined Hamiltonian [ Table V] [Figs. 8(f-h)]. The determined values of the exchange interactions (Table V) indicate a coupled spin-dimer model where the strongest exchange interaction ($J_2$) is between the 7$^{th}$ NN Cu ions having a distance of 5.125(3) Å, as proposed by Janson *et al*. [11]. The Cu spin-dimers are coupled within the *bc* plane by three exchange interactions with comparable strengths $J_1 = 2.8 \pm 0.03$ meV (AFM), $J_3 = 3.61 \pm 0.01$ meV (AFM), and $J_4 = 3.23 \pm 0.05$ meV (AFM). Such magnetic *bc* planes are weakly coupled by the exchange interaction $J_5 = 0.03 \pm 0.01$ meV. This experimentally determined exchange hierarchy (2D coupled spin-dimer model) is consistent with LDA-based density functional theory calculations by Janson *et al*. [11] and in sharp contrasts with predicted zig-zag spin-chain model by Yang *et al*. using GGA+U-based calculations [28]. Although the experimentally determined spin-model matches with that of the theoretically predicted model by Janson *et al*. [11] with ~ 90% agreement for the in-plane exchange interactions strengths, the strength of the interplanar coupling $J_5$ is experimentally found to be much weaker (~ 30% of the theoretical predicted value). In addition, we mention here that the earlier calculations do not consider the weak but finite single-ion-anisotropy. We have found that the single-ion-anisotropy term $D$ is essential along with the interplanar coupling $J_5$ to reproduce the experimental INS spectra, especially, the energy gap values. Moreover, the observed field-induced metamagnetic transition confirms the presence of single-ion-anisotropy.

To gain further insights of the spin-Hamiltonian, we have simulated single crystal spin-wave dispersions (using the determined values of $J_1$, $J_2$, $J_3$, $J_4$, $J_5$, and $D$) for the different principle reciprocal directions [Fig. 9(a)]. There are total of 8 spin-wave dispersion modes with two antiphase branches having 4-modes each. The dispersion curves with respective intensities are shown in Fig. 9(b). The intensities of dispersion modes modulate strongly with the momentum transfer and show strong intensity at and around the AFM zone centres (0,1,0) and (1,1,1). The dispersion modes are degenerate except around the AFM zone centres where the four modes (M1, M2, M3, and M4) have different energy gaps [0.80, 1.17, 1.08 and 1.37 meV] for (0,1,0) and [1.17, 0.80, 1.37 and 1.08 meV] (1,1,1). The detailed plots of the dispersions at around both the AFM zone centres (0,1,0) and (1,1,1) over 0.5–2 meV [Figs. 9(c) and 9(d)], respectively reveal unambiguously four modes (M1, M2, M3, and M4). At the AFM zone centre (0,1,0), only the M2 and M4 modes (among the four) have intensity, and have higher energy gaps of 1.17 and 1.37 meV, respectively. On the other AFM zone centre (1,1,1), the same M2 and M4 modes have intensity, however, they have lower energy gaps of 0.80 and 1.08 meV, respectively. Such variations of the energy gaps of the modes along with respective intensity result into two different observable energy gap values in the experimental powder INS spectra [Figs. 8(e) and 8 (g-h)]. We have found that the both $J_5$ and single-ion anisotropy parameter $D$ are responsible for the removal of the degeneracy of the dispersion modes (M1, M2, M3, and M4) at the AFM zone centres. However, the effects of $J_5$ and $D$ are contrastingly different at the AFM zone centres (0,1,0) and (1,1,1) as described below.

The importance of a small, non-zero $J_5$ is shown in Fig. 10. With the absence of the exchange interaction $J_5$ (i.e., $J_5$ = 0 meV) [values of all other parameters are kept fixed as given in Table V], only two modes are observable with energy gaps of 0.80 (1.08) meV, i.e., M1 and M2 (M3 and M4) are degenerate. The corresponding energy gap values are same at both the AFM zone centres (0,1,0) and (1,1,1). With a finite value of $J_5$, the degeneracies of the dispersion modes are removed. At the AFM zone centres (0,1,0), with the increase of the strength of $J_5$ exchange interaction, the energy gap values for M2 and M4 dispersion modes (having intensity) increases, whereas, the energy gap values for M1 and M3 dispersion modes (having no intensity) remains almost constant. In sharp contrast, an opposite scenario has been found for the other AFM zone centre (1,1,1). Where, the energy gap values for M2 and M4 dispersion modes (having intensity) remain almost constant, and the energy gap values for M1 and M3 dispersion modes (having no intensity) increases with the increasing strength of the exchange interaction $J_5$. Such opposite behaviours of the dispersion modes with $J_5$ at the AFM zone centres (0,1,0) and (1,1,1) lead to the observation of the different energy gaps of 1.17 and 0.80 meV, respectively, in the experimental spectra. The detailed variations of energy gaps for all the M1-M4 dispersion modes with the strength of $J_5$ are shown in Figs. 11(a) and 11(b). Around $J_5$ = 0.02 meV, the gap value of M2 dispersion mode becomes higher than that of the M3 dispersion mode, reflecting a band crossing.

Now we discuss the effect of single-ion anisotropy $D$ on the low energy dispersion modes [Fig. 12]. For $D$ = 0 [keeping the values of all other parameters fixed as given in Table V], again only two modes are observable, revealing the doubly degenerate modes. For $D$ = 0, the modes M1 and M3 (M2 and M4) are found to be degenerate. In this case, for the AFM zone centre (0,1,0), there is no energy gap for M1 and M3 dispersion modes (having no intensity), whereas, an energy gap of 0.85 meV is present for M2 and M4 dispersion modes (having intensity). In contrast, an opposite scenario has been found for the other AFM zone centre (1,1,1) where the M2 and M4 dispersion modes (having intensity) have no energy gap and the M1 and M3 dispersion modes (having no intensity) have an energy gap of 0.85 meV. With a finite $D$ value, the degeneracies of the dispersion modes are removed as well as an overall energy gap appears for both the AFM zone centres (0,1,0) and (1,1,1). With the increase of $D$ value, overall increase of the energy gaps for all four dispersion modes M1-M4 have been found for both the AFM zone centres (0,1,0) and (1,1,1) [Figs. 11(c) and 11 (d)]. The rate of increase of the energy gap values for lower energy modes [i.e., {M1 and M3 for (0,1,0)} and {M2 and M4 for (1,1,1)}] are higher than that for the higher energy modes [i.e., {M2 and M4 for (0,1,0)} and {M1 and M3 for (1,1,1)}] [Figs. 11(c) and 11 (d)]. Nevertheless, the effect of the $D$ value is same for both the AFM zone centres (0,1,0) and (1,1,1) [Figs. 11(c) and 11 (d)]. In addition, a band crossing is also evident with the increasing $D$ value [Figs. 11(c) and 11 (d)] similar to the case of $J_5$.

The above analyses unambiguously reveal that both interplanar exchange interaction $J_5$ and single-ion anisotropy $D$ are present in the studied compound $\alpha$-$Cu_2P_2O_7$. The experimentally observed gap values of 1.17 and 0.80 meV are only reproduced by a combination of the $J_5$ = 0.03 meV and $D$ = - 0.07 meV. We would like to mention here that in the absence of both $J_5$ and $D$ (i.e, $J_5$ = 0 and $D$ = 0), all the four dispersive modes become degenerate and the energy gaps vanishes at all the AFM zone centres [Figs. 11(e) and 11(f)].

Now we present further analyses to distinguish between theoretically proposed two spin-dimer models for $\alpha$-$Cu_2P_2O_7$, i.e., (i) the determined spin-dimers model (Model-1) constituted by the 7$^{th}$ NN Cu spins separated by larger distance [$d_{Cu\text{-}Cu}$ = 5.125(3) Å] and (ii) spin-dimers model (Model-2) formed by the NN Cu spins separated by the shortest distance [$d_{Cu\text{-}Cu}$ = 3.014(1) Å] [Fig. 13]. For this purpose, we have simulated additional excitation spectra considering an alternating values of $J_1$ and $J_2$; i.e., $J_1$ = 7.73 meV (AFM) and $J_2$ = 2.8 meV (AFM), and keeping the values of other parameters unchanged [Figs. 13(c) and 13(f)]. For a comparison, the simulated spectra for Model-1 are also included in Figs. 13 (b) and 13(e). The simulated single crystal excitation spectra [Figs. 13(e) and 13(f)] for the two models (Model-1 and Model-2) look almost similar in terms of dispersion band widths and energy gaps, except few differences in intensity around the zone boundaries (over ~ 9-12.5 meV) [highlighted with the dashed rectangles]. The Model-2 reveals higher intensities around the zone boundaries (over ~ 9-12.5 meV) as compared to that of the Model-1. Such differences in the intensity leads to an observable difference in the intensity of the simulated powder spectra over the energy range of 9-10.5 meV and $|Q|$ range of 0.9-2.1 Å$^{-1}$ for Model-1 and Model-2 [Figs. 13 (b) and 13(c)]. The measured experimental spectrum with $E_i$ = 28.2 meV and at 7 K is shown in Fig. 13(a) for a comparison. It is evident that the spectrum for Model-1 is in better agreement than that of the Model-2. For a quantitative comparison, we have shown the experimental constant-energy $|Q|$-cut plot in Fig. 13(d) along with the simulated curves for Model-1 and Model-2. It is evident that Model-1 provides better agreement with the experimental curve than that of the Model-2. Further comparisons of Model-2 with experimental spectra have been performed by tuning the values of in-pane inter-dimer exchange interactions $J_3$ and $J_4$ (for details see Appendix and Fig. 14). However, it is evident that Model-2 does not provide a better agreement with the experimental curve even with the adjustments of $J_3$ and $J_4$ values. Therefore, we conclude that the Model-1, i.e., the spin-dimers model constituted by the 7th NN Cu spins separated by larger distance [$d_{Cu\text{-}Cu}$ = 5.125(3) Å] is more appropriate for the studied compound $\alpha$-$Cu_2P_2O_7$ in agreement with the predicted model by Janson *et al*., [11].

## IV. DISCUSSION

Our in-depth combined INS measurements and spin-wave analyses establish that the magnetic model for the quantum magnet $\alpha$-$Cu_2P_2O_7$ is a two-dimensionally coupled spin-dimer system within the *bc* plane, rather than a simple nearest-neighbor coupled network. The magnetic dimers are formed between 7th NN Cu–Cu ions, despite their relatively large separation, with a dominant intradimer exchange interaction $J_2$ (7.73 ± 0.02 meV) and sizeable interdimer couplings $J_1$ (2.8 ± 0.03 meV), $J_3$ (3.61 ± 0.01 meV), and $J_4$ (3.23 ± 0.05 meV); and a weaker interplanar coupling $J_5$ (0.03 ± 0.01 meV) responsible for the emergence of long-range AFM order state below $T_N \approx 25$ K. The strongest exchange interaction, $J_2$, occurs between the 7th NN Cu-ion pair having a distance $d_{Cu\text{-}Cu}$ = 5.125(3) Å, in contrast to the weaker exchange interactions between the NN [$d_{Cu\text{-}Cu}$ = 3.014(1) Å], and 2nd -6th NN [$d_{Cu\text{-}Cu}$ =3.363(3), 3.855(3), 4.363(2), 4.812(3), and 4.840(1) Å] (out-of *bc*-plane) Cu-ions pairs. This shows that the magnitudes of the exchange interactions are not determined by the Cu···Cu distances, rather by the critical orientation and overlap of the active orbitals with the intervening ligands. It was reported that the magnetism of $\alpha$-$Cu_2P_2O_7$ originate from the half-filled Cu $3d_{x^2-y^2}$ orbital states and the strongest exchange interaction $J_2$ occurs through the Cu–O3–O4–Cu exchange pathways [11]. It was also reported that for the NN [$d_{Cu-Cu}$ =3.014(1) Å] Cu-Cu ion pair, the Cu-O-Cu angle is close to 90°, resulting into a sizable competing ferromagnetic contribution which leads to an effective weak AFM exchange interaction strength for $J_1$. The 2nd-6th NN [$d_{Cu-Cu}$ =3.363(3), 3.855(3), 4.363(2), 4.812(3), and 4.840(1) Å] Cu-Cu ion-pairs are between the *bc*-planes where the sizable FM contributions significantly reduce the strengths of the exchange couplings, effectively making them two orders of magnitude weaker than that of the in-plane exchange interactions. On the other hand, the FM contribution to the exchange interaction $J_2$ for the 7th NN Cu-ion pair [$d_{Cu-Cu}$ =5.125(3) Å] through the Cu–O3–O4–Cu pathway is minimal which results into the strongest AFM exchange interaction (7.73 ± 0.02 meV). The 8th [$d_{Cu-Cu}$ = 6.212(1) Å] and 9th [$d_{Cu-Cu}$ = 6.255(3) Å] NN Cu-Cu ion pairs form in-plane interdimer exchange interactions $J_3$ and $J_4$ through the Cu–O2–O4–Cu and Cu–O2–O3–Cu, respectively, where the FM contributions are small leading to sizeable strengths of AFM exchange interactions; $J_3$ =3.61 and $J_4$ =3.23 meV, respectively.

In the studied compound $\alpha$-$Cu_2P_2O_7$, the dominant intradimer exchange interaction is determined to be $J_2$ = 7.73 meV (this work), corresponding to the Cu-ion pair with separation of $d_{Cu-Cu}$ = 5.125(3) Å. For this Cu-ion pair, associated long-rage superexchange pathway cannot be assessed in terms of simple empirical considerations, such as the Goodenough-Kanamori-Anderson rule. Moreover, for isostructural compound $\alpha$-$Cu_2As_2O_7$ [10] as well, the reported strongest $J_2$ value is ≈ 7.78 meV [$d_{Cu-Cu}$ =5.239 Å]. In these cases, it was reported that the *p* orbitals of cation P/As do not hybridize strongly with the oxygen orbitals, and favourable superexchange takes place through the Cu–O–O–Cu pathway [11]. For the NN exchange interaction $J_1$, the strength of AFM exchange interaction varies with Cu–O–Cu bond angles, vis., the weakest interaction $J_1$ ≈ 0.43 meV occurs for $\beta$-$Cu_2V_2O_7$ having smallest bond-angle Cu–O–Cu ≈ 98.7° [11,12], a relatively stronger $J_1$ ≈ 2.8 meV (this work) is found for $\alpha$-$Cu_2P_2O_7$ having larger bond-angle Cu–O–Cu ≈ 99.9°, and strongest exchange interaction $J_1$ ≈ 14.14 meV is reported for $\alpha$-$Cu_2As_2O_7$ having the largest bond-angle Cu–O–Cu ≈ 101.7° [10]. This indicates that larger bond angles (deviation from 90°) favour stronger AFM NN exchange interactions. The similar magnetic exchange interaction topology within the *bc* plane in $\alpha$-$Cu_2As_2O_7$ leads to an identical magnetic structure as like $\alpha$-$Cu_2P_2O_7$ with magnetic space group *C2/c′* [10,11]. On the other hand, the different nature of the interdimer couplings within the *bc* plane of $\beta$-$Cu_2V_2O_7$ results into a distinct magnetic structure with magnetic space group *C2′/c* [54].

Now, we discuss the role of the finite but weak interplanar exchange interaction $J_5$ (0.03± 0.01 meV). In the present compound $\alpha$-$Cu_2P_2O_7$, the spin-dimers are strongly coupled within the *bc* plane through the dominant intraplanar exchange interactions ($J_1$, $J_2$, $J_3$ and $J_4$), giving raise the robust two-dimensional antiferromagnetic correlations. As per Mermin-Wagner theorem, a pure two-dimensional isotropic Heisenberg system cannot sustain a long-range magnetic order at finite temperature because of thermal and quantum fluctuations. The interplanar exchange interaction $J_5$ provides the three-dimensional magnetic connectivity between the *bc* planes in $\alpha$-$Cu_2P_2O_7$ and stabilized the observed long range AFM order below the $T_N$, despite its small magnitude; ~ 100 times weaker as compared to the intraplanar couplings. The behavior is consistent with many quasi-two-dimensional antiferromagnets, where long-range order emerges through weak but finite interplanar coupling that bridges strongly-correlated magnetic layers. For examples, two-dimensional triangular-lattice $Na_3Fe(PO_4)_2$ exhibits long-range magnetic ordering below $T_N$ = 10.8 K even though the interplanar exchange interaction is ~ 400 times weaker than the dominant intraplanar exchange interactions [55]. Similarly, long-range magnetic order is observed in the kagome-lattice $Cs_2Cu_3SnF_{12}$ ($T_N$ = 20.2 K) [56], honeycomb-lattice $Na_2Ni_2TeO_6$ ($T_N$ = 27.5 K) [51], and triangular-lattice $Ba_3CoSb_2O_9$ ($T_N$ = 3.8 K) [57], where the interplanar exchange interaction is ~ 100 times weaker than the corresponding dominant intraplanar interactions.

Here, we compare the interplanar exchange coupling and anisotropy among the isostructural compounds. A sizable interplanar exchange interaction $J_\perp$ (1.47 meV) was reported in $\beta$-$Cu_2V_2O_7$ ($T_N \approx$ 26 K) [12,58], whereas no such interplanar exchange interaction (or assumed to be very weak) was reported for $\alpha$-$Cu_2As_2O_7$ ($T_N \approx$ 10 K) [10]. The exchange interaction $J_5$ in the studied compound $\alpha$-$Cu_2P_2O_7$ occurs through the Cu–O3–Cu [$d_{Cu-Cu}$ = 3.363(3) Å] superexchange pathway, whereas the exchange interaction $J_\perp$ in $\beta$-$Cu_2V_2O_7$ occurs through the $V_2O_7$ groups [$d_{Cu-Cu}$ = 7.32 Å]. It has been reported that a sizable FM contribution significantly reduces the strength of the interplanar exchange interaction $J_5$ in $\alpha$-$Cu_2P_2O_7$. Conversely, due to the weak FM contribution to $J_\perp$, it reveals sizable interplanar exchange interaction even though the Cu···Cu distance is considerably larger [$d_{Cu-Cu}$ = 7.32 Å] in $\beta$-$Cu_2V_2O_7$. In addition, present study determines a weak but finite *D* of - 0.07 meV, resembled with the energy gaps in the excitation spectrum. The observed field-induced metamagnetic transition at $H_C$ ~13 kOe (2 K), originating from the spin-reorientation transition within the AFM sublattices, may be associated with the single-ion anisotropy. In fact, slightly anisotropic *g*-factors of $g_\perp$ ~ 2.1 and $g_\parallel$ ~ 2.4 were reported from ESR study [11]. Metamagnetic transitions are also reported for other isostructural compounds, viz., for $\beta$-$Cu_2V_2O_7$ with $H_C$ ~15 kOe (5 K) [58] and for $\alpha$-$Cu_2As_2O_7$ with $H_C$ ~ 17 kOe (2 K) [10]. For these compounds as well slightly anisotropic g-factors were reported ($g_\perp$ ~ 2.09 and $g_\parallel$ ~ 2.23 for $\beta$-$Cu_2V_2O_7$ [59]; and $g_\perp$ ~ 2.18 and $g_\parallel$ ~ 2.07 for $\alpha$-$Cu_2As_2O_7$ [10].

Below we provide the microscopic origin for the preservation of the crystallographic centrosymmetry of the lattice under the observed AFM phase transition in $\alpha$-$Cu_2P_2O_7$. Retention of crystallographic symmetries such as centering, translation, inversion, mirror and glide planes in the magnetically ordered state is rather unusual in many antiferromagnetic compounds, which often lower symmetry upon magnetic ordering [60-62]. The compound $\alpha$-$Cu_2P_2O_7$ crystallizes in the *C2/c* space group which features the *C*-centering of the unit cell (non-primitive lattice) and the magnetic structure adopts the magnetic space group *C2/c′* (BNS notation, No. 15.88) with a propagation vector $\boldsymbol{k}$ = (0,0,0). The *C*-centering translation {1/2 1/2 0} for $\boldsymbol{k}$ = (0,0,0) yields a phase factor of $e^{i2\pi} = 1$ on the magnetic moments. Hence, the magnetic moment directions related by *C*-centering symmetry operation remain identical (parallel) at all Cu atom positions [Table III] in $\alpha$-$Cu_2P_2O_7$. Consequently, the magnetic unit cell remains *C*-centered. As a result, the antiferromagnetic order with $\boldsymbol{k}$ = (0,0,0) in $\alpha$-$Cu_2P_2O_7$ preserves the crystallographic *C*-centering symmetry, leading to a non-primitive magnetic unit cell described by the magnetic space group *C2/c′*. For the isostructural compounds $\beta$-$Cu_2V_2O_7$ and $\alpha$-$Cu_2As_2O_7$ as well, the magnetic space group contains *C*-centre symmetry. Similar preservation of *C*-centering has also been reported for other AFM compound $CaMnGe_2O_6$ (crystallizes in the monoclinic *C2/c* space group), where magnetic structure adopts the magnetic space group *C2′/c* (No. 15.87) with $\boldsymbol{k}$ = (0,0,0) [63]. For the materials with monoclinic space group *C2/c*, to break the *C*-centering symmetry upon a magnetic ordering, the propagation vector needs to be $\boldsymbol{k} \neq (0,0,0)$. For example, in AFM compound $SrCoGe_2O_6$, which crystallizes in the monoclinic *C2/c* space group, the magnetic propagation vector $\boldsymbol{k}$ = (1,0,0) introduces a phase factor of π (-1) under the *C*-centering translation, thereby breaking the *C*-centering symmetry and resulting in a primitive magnetic structure described by the magnetic space group $P_c2_1/c$ (No. 14.84) [64].

## SUMMARY AND CONCLUSION

In summary, we have carried out a comprehensive experimental and theoretical investigation of the magnetic properties of the low-dimensional quantum magnet $\alpha$-$Cu_2P_2O_7$. Using temperature-dependent inelastic neutron scattering, powder neutron diffraction, dc-magnetization measurements, and detailed spin-wave modeling, we have resolved the long-standing controversy regarding its microscopic magnetic Hamiltonian. Our results demonstrate that $\alpha$-$Cu_2P_2O_7$ is best described as a two-dimensionally coupled spin-dimerized antiferromagnetic lattice within the *bc* plane, rather than a simple nearest-neighbor coupled network. The magnetic dimers are formed between 7$^{th}$-nearest-neighbor Cu–Cu ions, despite their relatively large separation, with a dominant intradimer exchange interaction $J_2$ and weaker interdimer couplings $J_1$, $J_3$, and $J_4$. This exchange hierarchy is in excellent agreement with LDA-based density functional theory calculations and clearly contrasts with earlier GGA+U-based predictions. Weak interlayer coupling $J_5$ connects the two-dimensional magnetic layers into a three-dimensional lattice and stabilizes long-range antiferromagnetic order below $T_N \approx 25$ K. Our inelastic neutron scattering data reveal gapped spin-wave excitations, which are quantitatively reproduced by linear spin-wave theory only when a weak but finite single-ion anisotropy is included. This anisotropy originates from the distorted $CuO_5$ polyhedra and plays a crucial role in opening an excitation gap and driving a field-induced metamagnetic transition. The combined effects of interlayer coupling and anisotropy account for the presence of two distinct energy gaps observed at different antiferromagnetic zone centers. Temperature-dependent powder neutron diffraction measurements confirm the proposed magnetic structure and show that the antiferromagnetic ordering preserves the crystallographic centrosymmetry of the lattice. Furthermore, detailed field- and temperature-dependent magnetization measurements, supported by field-dependent neutron diffraction, uncover a previously overlooked metamagnetic transition near 13 kOe and enable the construction of a comprehensive magnetic phase diagram in the $H-T$ plane. Overall, our study establishes $\alpha$-$Cu_2P_2O_7$ as a rare realization of a two-dimensional spin-dimer system in which long-range magnetic order emerges from weak interlayer coupling and magnetic anisotropy. These findings highlight the importance of extended exchange pathways and subtle anisotropic interactions in governing the ground state and excitation spectrum of low-dimensional quantum magnets.

**Data Availability:**

The data that support the findings of this article are openly available [66]; embargo periods may apply.

## APPENDIX: EFFECT OF IN-PLANE INTER-DIMER EXCHANGE INTERACTIONS $J_3$ AND $J_4$ ON THE EXCITATION SPECTRA

It is shown earlier that Model-2 (i.e., the interchanging of the values of $J_1$ and $J_2$ values) does not provide good agreement with the experimental curve [Fig. 13(d)]. We have shown below the further tuning of the Model-2 by (i) interchanging the values of $J_3$ and $J_4$ (i.e., $J_3$ = 3.23 meV and $J_4$ = 3.61 meV); designated as Model-$2_1'$ as well as variations of $J_3$ and $J_4$ values i.e., (ii) considering a higher value of $J_3$ = 4.23 meV and a lower value of $J_4$ = 2.61 meV; designated as Model-$2_2'$, and (iii) considering a lower value of $J_3 = 2.23$ meV and a higher value of $J_4 = 4.61$ meV; designated as Model-$2'_3$. The simulated curves with the Models $2'_1$, $2'_2$, and $2'_3$ are compared with the experimental spectra in Fig. 14. In all these cases, simulated curves do not reproduce the experimental data as like Model-1. Therefore, neither Model-2 nor any adjustment of the values of $J_3$ and $J_4$ (either by interchanging or by tuning the values) in Model-2 reproduce the $Q$-dependent intensity of the experimental excitation spectrum [Fig. 14]. In contrast, as discussed earlier in the main text the Model-1 provides overall the best agreement with the experimental curve [Fig. 13(d)].

**Tables:**

TABLE I. The Rietveld-refined lattice parameters, fractional atomic coordinates, isotropic thermal parameters and site occupancy of $\alpha$-$Cu_2P_2O_7$ (*C2/c*), obtained from the analysis of neutron diffraction pattern, measured at room temperature. The lattice parameters and fractional atomic coordinates obtained from DFT calculation.

| Lattice parameter | | Experimental | | | | | Theoretical | | |
|---|---|---|---|---|---|---|---|---|---|
| *a* (Å) | | 6.883(1) | | | | | 6.876 | | |
| *b* (Å) | | 8.115(2) | | | | | 8.113 | | |
| *c* (Å) | | 9.166(3) | | | | | 9.162 | | |
| $\beta$ (deg.) | | 109.56(2) | | | | | 109.54 | | |
| Atom | Site | | | | | | | | |
| | | *x/a* | *y/b* | *z/c* | $B_{iso}$ | *Occ.* | *x/a* | *y/b* | *z/c* |
| Cu | 8*f* | 0.9851(8) | 0.3152(6) | 0.5085(8) | 0.36(2) | 1 | 0.9862 | 0.3097 | 0.5045 |
| P | 8*f* | 0.2022(8) | 0.0045(8) | 0.2074(7) | 0.35(2) | 1 | 0.2008 | 0.0065 | 0.2054 |
| O1 | 4*e* | 0 | 0.0500(1) | 0.25 | 0.94(4) | 1 | 0 | 0.0361 | 0.25 |
| O2 | 8*f* | 0.3734(9) | 0.9988(6) | 0.3629(6) | 0.58(3) | 1 | 0.3811 | 0.9991 | 0.3627 |
| O3 | 8*f* | 0.2196(9) | 0.1524(8) | 0.1117(8) | 0.37(3) | 1 | 0.2238 | 0.1602 | 0.1165 |
| O4 | 8*f* | 0.1734(6) | 0.8475(9) | 0.1160(8) | 0.33(5) | 1 | 0.1875 | 0.8494 | 0.1138 |

TABLE II. The bond lengths (Cu–O, P–O) and bond angles (O–Cu–O, O–P–O) within the $CuO_5$ and $PO_4$ polyhedra at 300 K.

| Polyhedra | Bond | Distance (Å) | Bond angle paths | Angle (deg.) |
|---|---|---|---|---|
| $CuO_5$ | Cu–O3 | 2.339 (2) | Within basal plane | |
| | Cu–O4 | 1.882(5) | O2–Cu–O2 | 80.1(3) |
| | Cu–O2 | 1.947(1) | O2–Cu–O3 | 95.1(4) |
| | Cu–O3 | 1.977(3) | O3–Cu–O4 | 93.5(4) |
| | Cu–O2 | 1.991(1) | O4–Cu–O2 | 95.9(4) |
| | | | Perpendicular to basal plane | |
| | | | O3–Cu–O4 | 111.3(5) |
| | | | O3–Cu–O3 | 78.1(4) |
| | | | O3–Cu–O2 | 90.4(4) |
| | | | O3–Cu–O2 | 88.3(4) |
| $PO_4$ | P–O1 | 1.611(7) | O1–P–O2 | 103.9(5) |
| | P–O2 | 1.516(7) | O2–P–O3 | 111.8(6) |
| | P–O3 | 1.515(5) | O3–P–O4 | 111.6(6) |
| | P–O4 | 1.501(6) | O4–P–O1 | 111.5(6) |
| | | | O3–P–O1 | 101.6(6) |
| | | | O2–P–O4 | 115.2(7) |

TABLE III. List of Wyckoff (8*f*) positions and possible components of magnetic moment (*m*) corresponding to Cu atom which are symmetry related of the maximal magnetic space groups (MSGs) [in Belov–Neronova–Smirnova (BNS) standard setting [65]] for the parent space group *C2/c* (No. 15) and the propagation vector ***k*** = (0,0,0) for the compound $\alpha$-$Cu_2P_2O_7$.

| **Wykoff Position** (Magnetic atom Cu) 8*f* | **Magnetic space group (MSG)** | | | |
|---|---|---|---|---|
| | *C2'/c'* (No. 15.89) | *C2/c'* (No. 15.88) | *C2'/c* (No. 15.87) | *C2/c* (No. 15.85) |
| ($x, y, z$) | ($m_x, m_y, m_z$) | ($m_x, m_y, m_z$) | ($m_x, m_y, m_z$) | ($m_x, m_y, m_z$) |
| ($-x, y, -z+1/2$) | ($m_x, -m_y, m_z$) | ($-m_x, m_y, -m_z$) | ($m_x, -m_y, m_z$) | ($-m_x, m_y, -m_z$) |
| ($-x, -y, -z$) | ($m_x, m_y, m_z$) | ($-m_x, -m_y, -m_z$) | ($-m_x, -m_y, -m_z$) | ($m_x, m_y, m_z$) |
| ($x, -y, z+1/2$) | ($m_x, -m_y, m_z$) | ($m_x, -m_y, m_z$) | ($-m_x, m_y, -m_z$) | ($-m_x, m_y, -m_z$) |
| ($x, y, z$)+{1/2 1/2 0} | ($m_x, m_y, m_z$) | ($m_x, m_y, m_z$) | ($m_x, m_y, m_z$) | ($m_x, m_y, m_z$) |
| ($-x, y, -z+1/2$)+{1/2 1/2 0} | ($m_x, -m_y, m_z$) | ($-m_x, m_y, -m_z$) | ($m_x, -m_y, m_z$) | ($-m_x, m_y, -m_z$) |
| ($-x$ $-y, -z$)+{1/2 1/2 0} | ($m_x, m_y, m_z$) | ($-m_x, -m_y, -m_z$) | ($-m_x, -m_y, -m_z$) | ($m_x, m_y, m_z$) |
| ($x, -y, z+1/2$)+{1/2 1/2 0} | ($m_x, -m_y, m_z$) | ($m_x, -m_y, m_z$) | ($-m_x, m_y, -m_z$) | ($-m_x, m_y, -m_z$) |

TABLE IV. Total energies (relative to the lowest-energy configuration) per unit cell, and average Cu magnetic moments for the four magnetic configurations of $\alpha$-$Cu_2P_2O_7$ calculated by GGA and GGA+U methods using VASP software package.

| Magnetic space group (MSG) | GGA | | GGA+U | |
|---|---|---|---|---|
| | ΔE (meV/f.u.) | $\langle m_{Cu} \rangle$ ($\mu_B/Cu^{2+}$) | ΔE (eV/f.u.) | $\langle m_{Cu} \rangle$ ($\mu_B/Cu^{2+}$) |
| *C2'/c'*(No. 15.89) | 336.501 | 0.641 | 38.52658 | 0.840 |
| *C2/c'* (No. 15.88) | 0 | 0.605 | 0 | 0.835 |
| *C2'/c* (No. 15.87) | 55.387 | 0.604 | 10.04351 | 0.835 |
| *C2/c* (No. 15.85) | 217.861 | 0.635 | 24.16304 | 0.837 |

TABLE V. A comparison of exchange interaction values obtained from the inelastic neutron scattering (INS) study and DFT calculations.

| Exchange interactions | Superexchange pathways and Cu····Cu direct distance (Å) | Value (INS)(meV) | Value DFT calculations [11] (meV) |
|---|---|---|---|
| $J_1$ | Cu–O2–Cu (in-plane)<br>Cu····Cu = 3.014(1) | 2.8 ± 0.03 | 2.93 |
| $J_2$ | Cu–O3–O4–Cu (in-plane)<br>Cu····Cu = 5.125(3) | 7.73 ± 0.02 | 8.79 |
| $J_3$ | Cu–O2–O4–Cu (in-plane)<br>Cu····Cu = 6.212(1) | 3.61 ± 0.01 | 3.96 |
| $J_4$ | Cu–O2–O3–Cu (in-plane)<br>Cu····Cu = 6.255(3) | 3.23 ± 0.05 | 3.53 |
| $J_5$ | Cu–O3–Cu (out of plane)<br>Cu····Cu = 3.363(3) | 0.03 ± 0.01 | 0.09 |
| $D$ | | -0.07 ± 0.02 | |

## Figures:

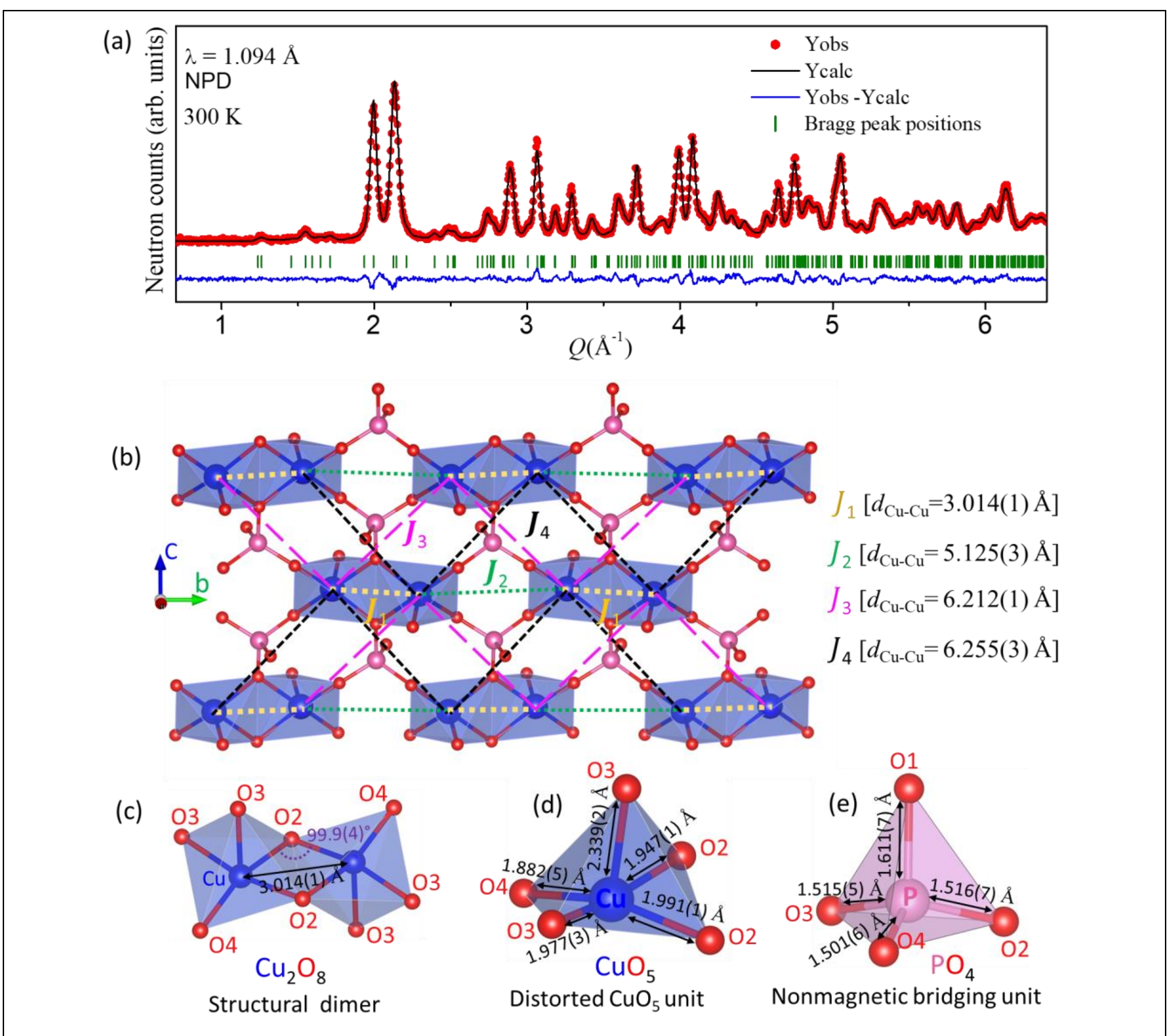

FIG. 1. (a) Rietveld analyzed neutron powder diffraction pattern of $\alpha$-$Cu_2P_2O_7$, measured at 300 K using the neutron diffractometer PD-1 (λ= 1.094 Å), BARC, India. The observed and calculated diffraction patterns are shown by the filled circles (red) and solid line (black), respectively. The difference between observed and calculated patterns is shown by the thin blue line at the bottom. The vertical bars are the allowed Bragg peak positions. (b) The representative Cu-Cu dimer crystal structure and possible exchange interactions within the *bc* plane. Individual exchange interactions are labelled by $J_1$ (yellow), $J_2$ (green), $J_3$ (pink), and $J_4$ (black). The magnetic layers, in the *bc* plane, are coupled by exchange interaction $J_5$ (not shown here) along the *a*-axis. The Cu, P and O ions are denoted by blue, pink and red spheres, respectively. (c) The details of the structural dimer ($Cu_2O_8$), formed by two edge-shared distorted $CuO_5$ square pyramids. (d) and (e) the local environments of $CuO_5$ square pyramid and $PO_4$ tetrahedra, respectively.

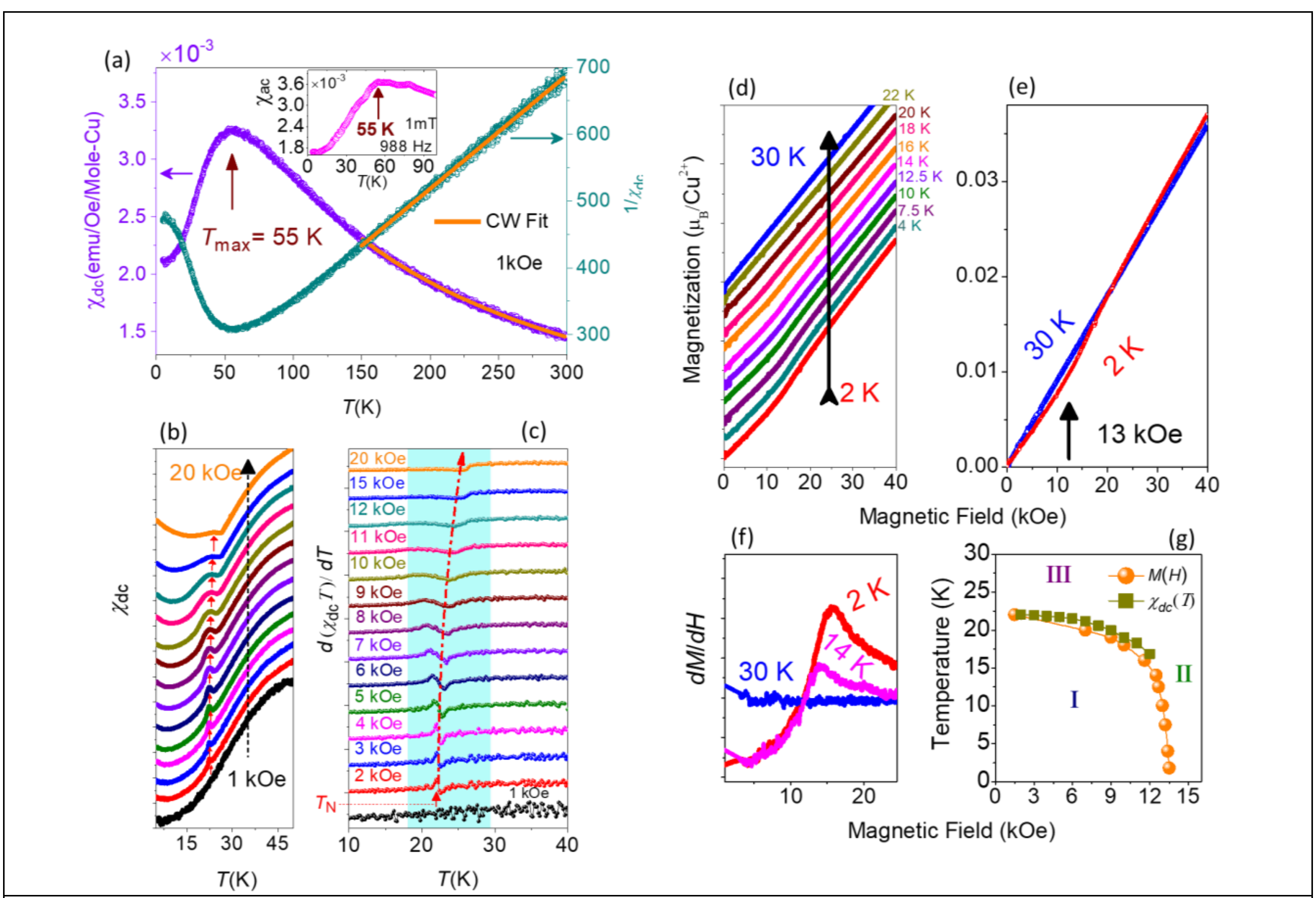

FIG. 2. (a) The temperature-dependent of dc-susceptibility ($\chi_{dc}$) and inverse of dc-susceptibility ($1/\chi_{dc}$) curves, measured under an applied field of 1 kOe, for $\alpha$-$Cu_2P_2O_7$. The solid lines through the data points are the fitted curves by Curie-Weiss (CW) law. The inset shows the temperature-dependent ac-susceptibility ($\chi_{ac}$) curve measured with 1mT ac-magnetic field with a frequency 988 Hz. (b) Temperature-dependent dc-susceptibility ($\chi_{dc}$) curves under the various applied magnetic fields between 1-20 kOe. For clarity, the curves are shifted vertically by a constant amount. (c) The temperature dependent Fisher's specific heat [$d(\chi_{dc}T)/dT$] curves under the various applied magnetic fields. For clarity, the curves are shifted vertically by a constant amount. (d) The field-dependent isothermal magnetization ($M$) curves of $\alpha$-$Cu_2P_2O_7$ measured at 2, 4, 7.5, 10, 12.5, 14, 16, 18, 20, 22 and 30 K. For clarity, the curves are shifted vertically by a constant amount. (e) The isothermal magnetization curves at 2 and 30 K, revealing the field -induced metamagnetic transition at a critical field of $H_C$ ~13 kOe at 2 K. (f) The field-dependent derivative of isothermal magnetization ($dM/dH$) at 2, 14 and 30 K, revealing the temperature dependence of the field-induced metamagnetic transition. (g) Magnetic phase diagram of $\alpha$-$Cu_2P_2O_7$ in the $H$-$T$ plane, obtained from the temperature and magnetic field dependent magnetization data, revealing three regions: (I) an AFM ordered state below ~13 kOe, (II) a field-induced metamagnetic phase above ~13 kOe, and (III) a short-range ordered (SRO) state above ~25 K.

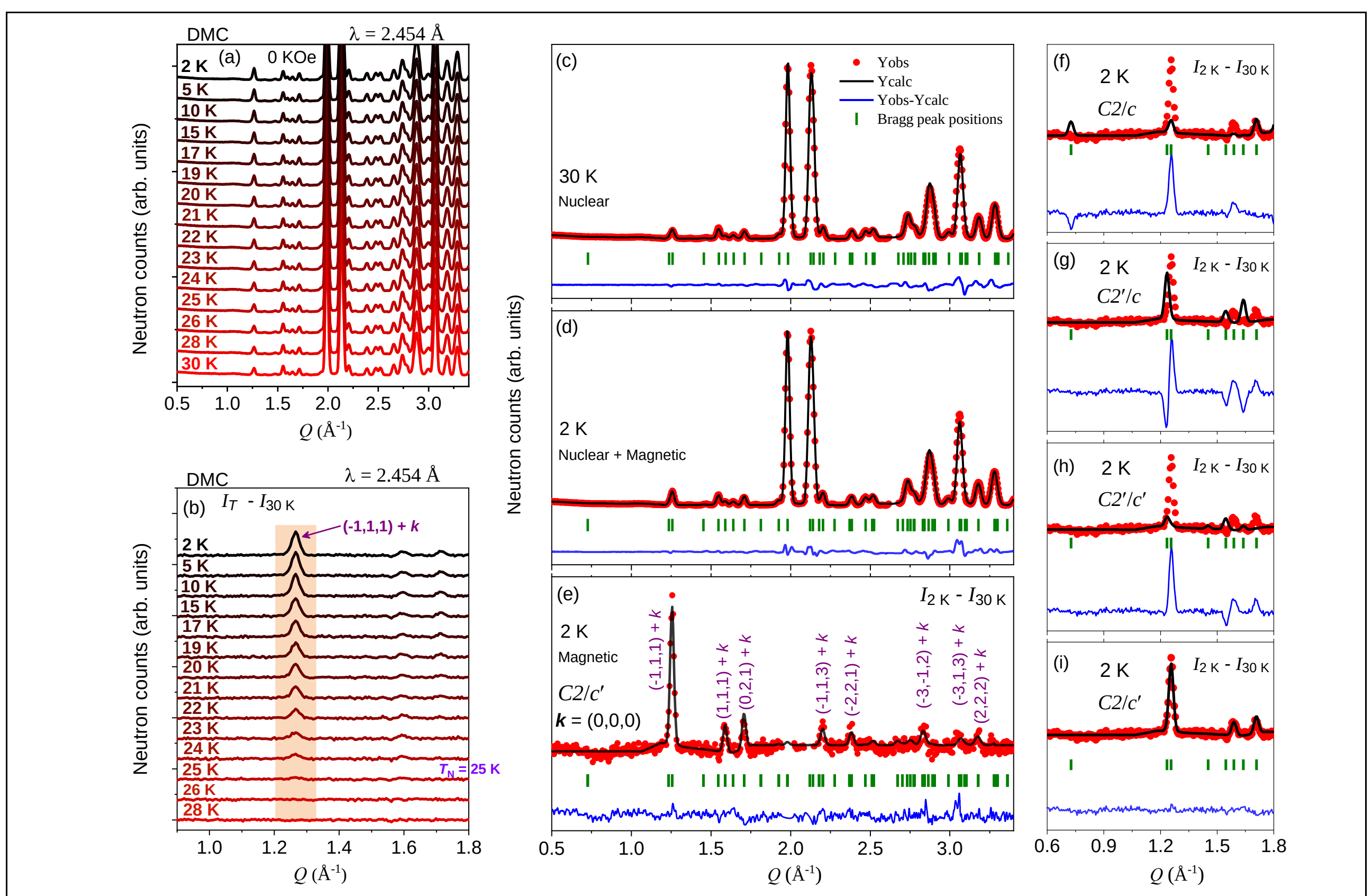

FIG. 3. (a) The temperature-dependent neutron diffraction patterns of $\alpha$-$Cu_2P_2O_7$ measured over the temperature range 2−30 K using the neutron diffractometer DMC (λ= 2.4586 Å), PSI, Switzerland. (b) The temperature-dependent pure magnetic diffraction patterns (after subtraction of nuclear background at 30 K) over the $Q$ = 0.9−1.8 Å$^{-1}$, highlighting the temperature-evaluation of strongest magnetic Bragg peak (-1, 1, 1 +***k***) and the magnetic ordering below $T_N \approx$ 25 K. Experimental and calculated neutron diffraction pattern at (c) 30 K (nuclear phase at paramagnetic state), and (d) 2 K [nuclear + magnetic phases (magnetic space group *C2*/*c*') at magnetically ordered state]. (e) Pure magnetic diffraction pattern at 2 K (after subtraction of nuclear background at 30 K) and calculated pattern by considering the magnetic phase (magnetic space group *C2*/*c*') alone. (f), (g), (h) and (i) The comparison of the experimentally measured pure magnetic diffraction pattern (2 K-30 K) with the calculated patterns by magnetic phase considering the magnetic space group *C2/c (No*. 15.85), *C2*'/*c* (No. 15.87), *C2*'/*c*' (No. 15.89) and *C2*/*c*' (No. 15.88), respectively.

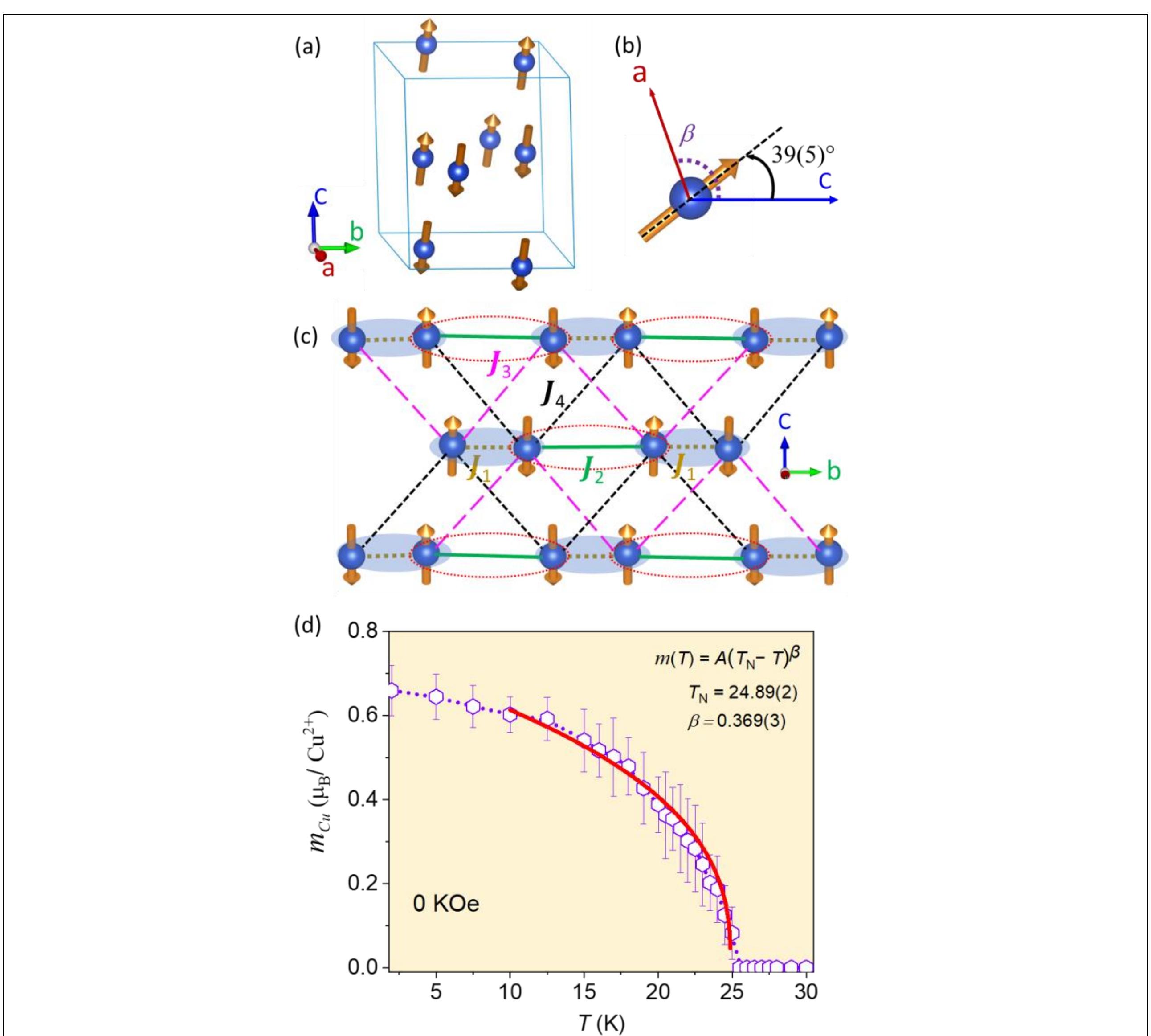

FIG. 4. (a) The schematic magnetic unit cell of $\alpha$-$Cu_2P_2O_7$. (b) The spin direction within the *ac* plane. (c) The magnetic structure within the *bc* plane along with the dimer models (structural dimes: shaded ovals and magnetic dimers: open ovals) having exchange interactions $J_1$, $J_2$, $J_3$, and $J_4$. (d) Temperature-dependent ordered magnetic moment of $Cu^{2+}$ ($m_{cu}$) for $\alpha$-$Cu_2P_2O_7$. The solid line (red) is the power-law [$m(T) = A(T_N - T)^{\beta}$] fit to the data.

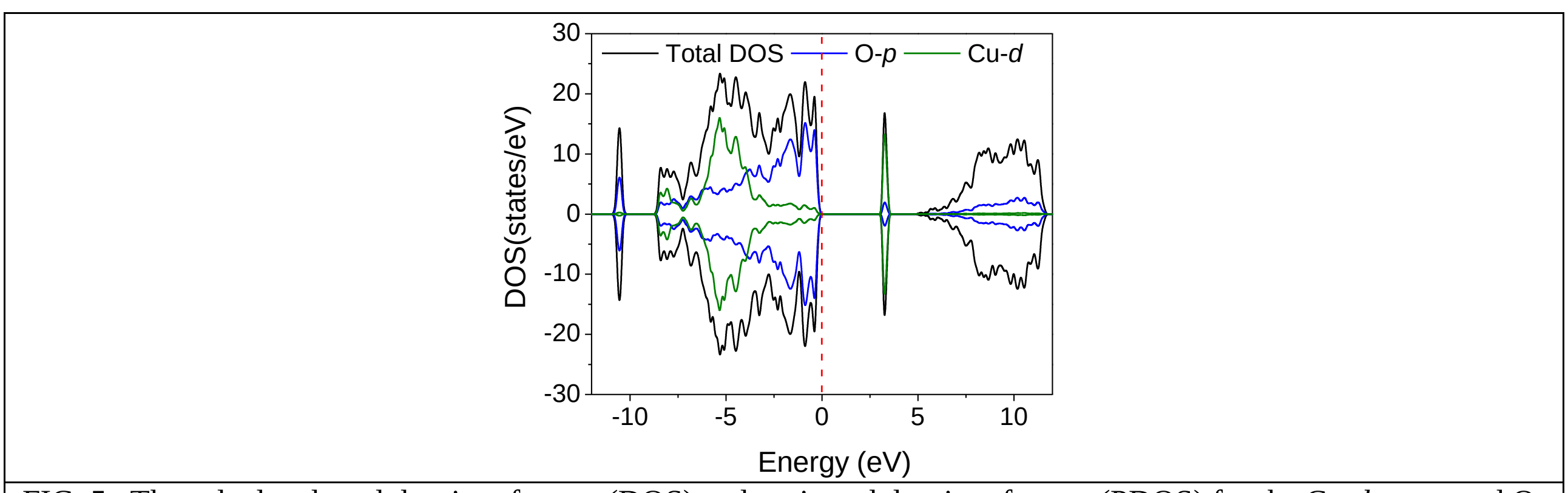

FIG. 5. The calculated total density of states (DOS) and projected density of states (PDOS) for the Cu-*d* states and O-*p* states calculated within GGA + U framework. The upper and lower panels represent spin-up and spin-down states, respectively. The black, green, and blue colors represent the total DOS, Cu-*d* states, and O-*p* states, respectively. Fermi energy is set to zero and shown by dashed vertical line (red).

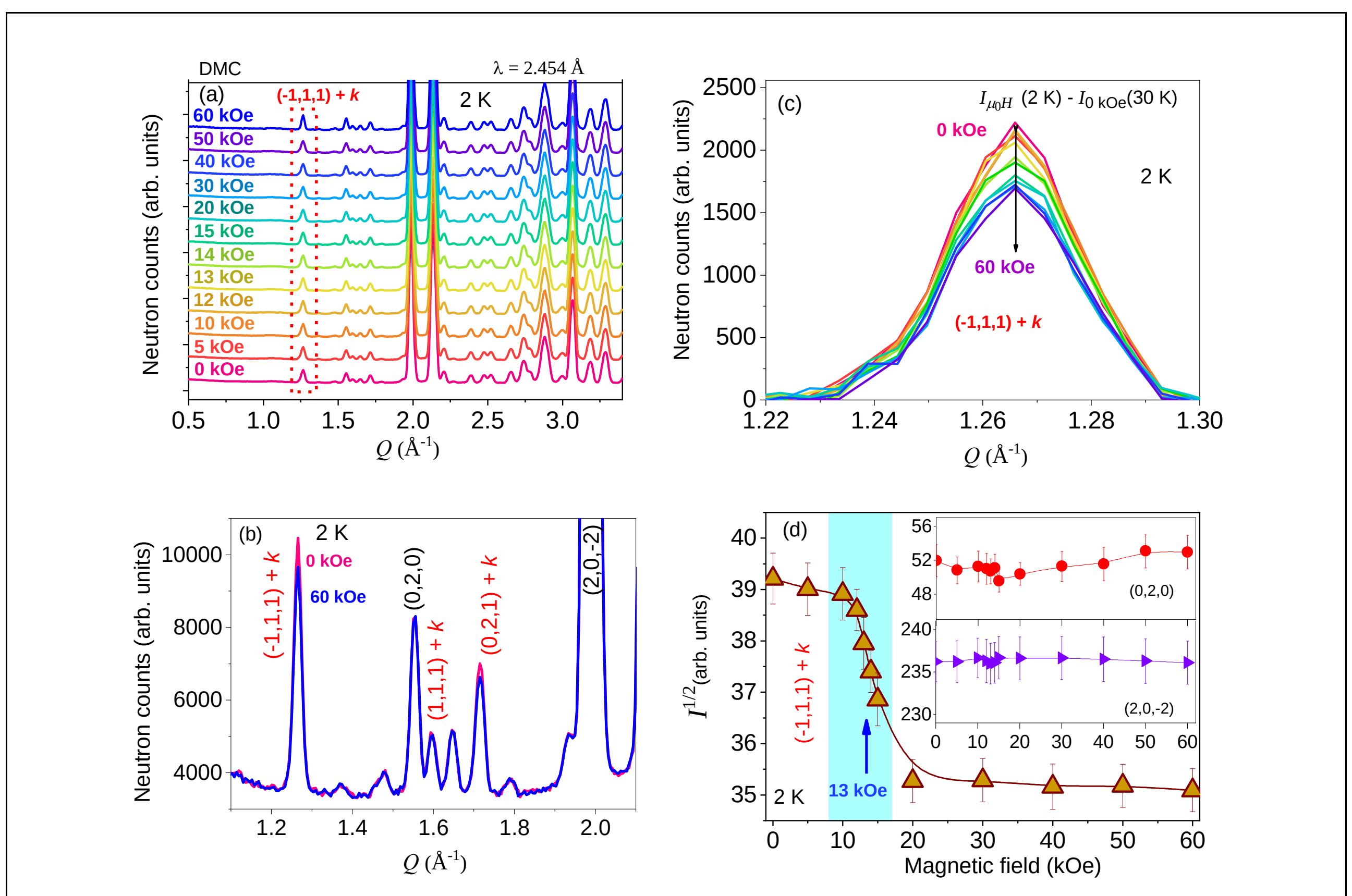

FIG. 6. (a) The field-dependent neutron diffraction patterns measured at 2 K on a pressed cylindrical-rod of powder samples of $\alpha$-$Cu_2P_2O_7$ using the cryo-magnet MA6, DMC (λ= 2.4586 Å), PSI, Switzerland. The patterns are vertically shifted by a constant amount for presentation purpose. (b) The direct comparison of patterns measured under 0 and 60 kOe magnetic fields. (c) The field-dependence of the strongest AFM Bragg peak (-1,1,1)+***k***, highlighting the variation of magnetic intensity with applied magnetic field. (d) The field variation of the square root of the integrated intensity of the strongest AFM Bragg peak (-1,1,1)+***k***. The insets show the field variations of the integrated intensities for the nuclear Bragg peaks (0,2,0) and (2,0-2).

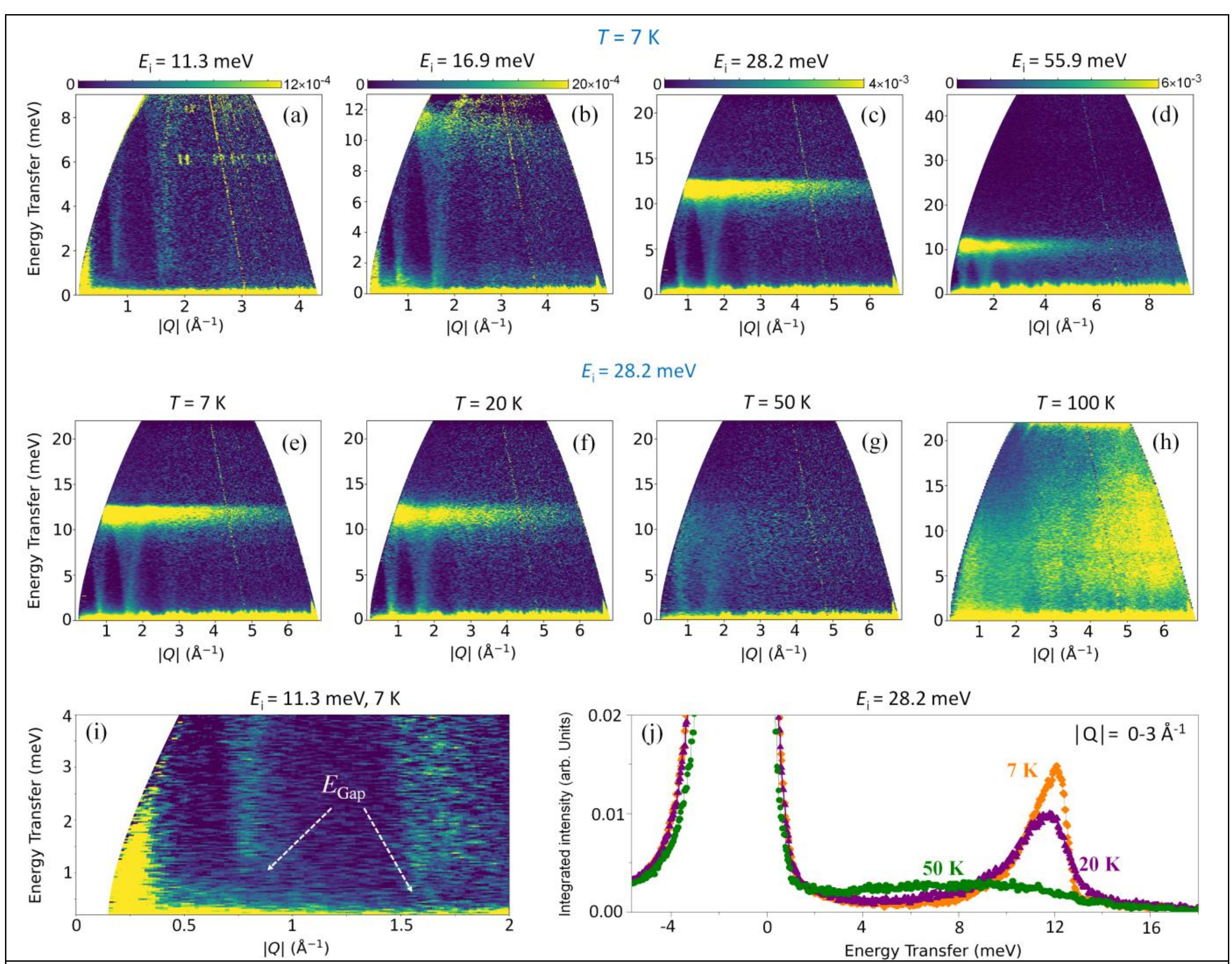

FIG. 7. 2D color maps of the INS spectra, measured on the powder samples of $\alpha$-$Cu_2P_2O_7$ using time-of-flight neutron spectrometer MERLIN, with incident energy ($E_i$) of (a) 11.3 meV, (b) 16.9 meV, (c) 28.2 meV, and (d) 55.9 meV. The color scales show the scattering intensity $S(|Q|,\omega)$ in arbitrary units. (e-h) Temperature dependent INS spectra measured at 7, 20, 50 and 100 K with $E_i$ = 28.2 meV. (i) The selected region of the INS spectrum measured at 7 K with $E_i$ = 11.3 meV over the energy range 0-4 meV, revealing the two distinct energy gaps of ~ 1.17 and 0.80 meV at $|Q|$ = 0.78 and 1.62 Å$^{-1}$, respectively. (j) The intensity vs energy curves obtained by integration over $|Q|$ = 0–3 Å$^{-1}$ of the INS spectra measured with $E_i$ = 28.2 meV.

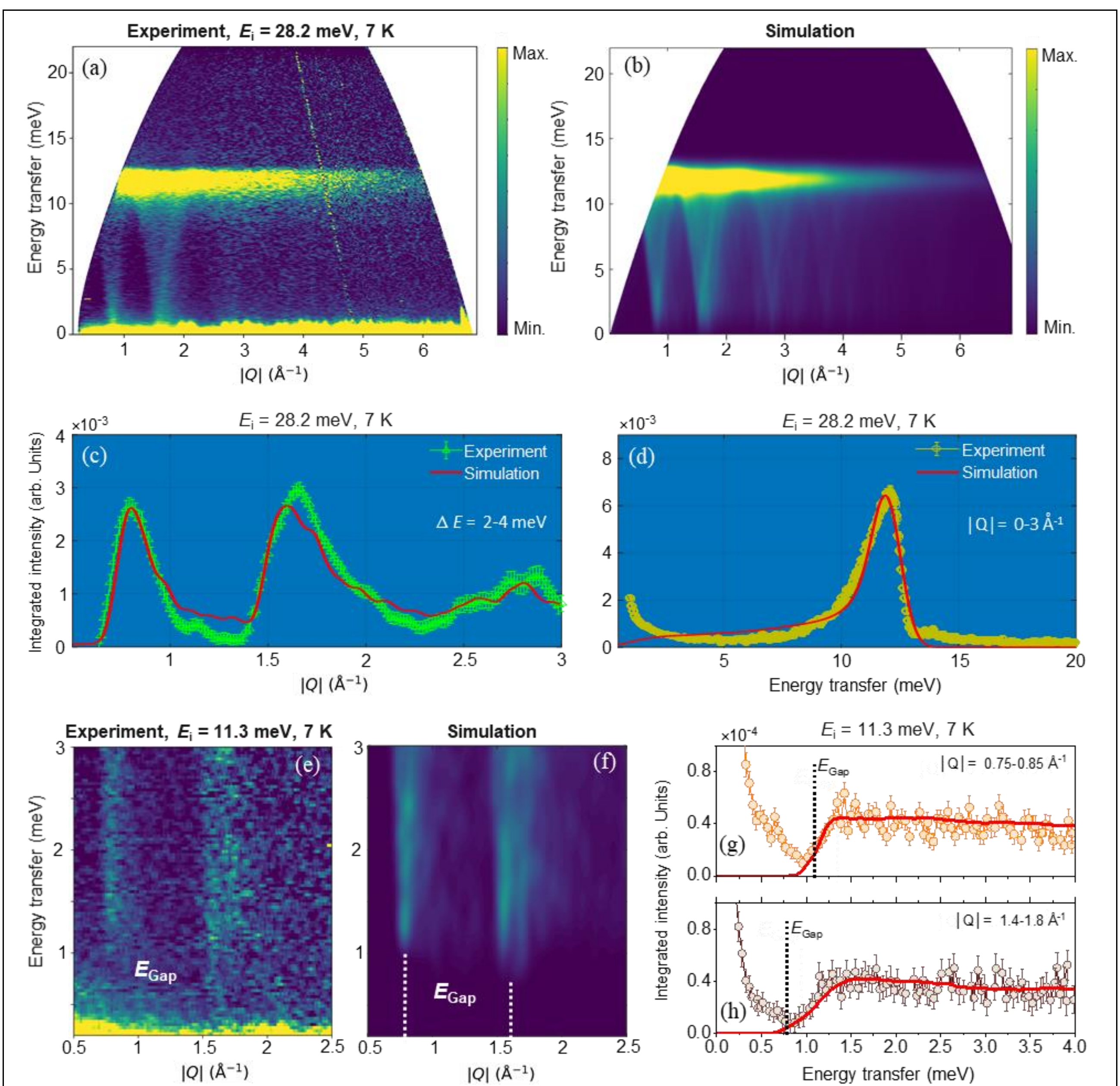

FIG. 8. (a) Experimentally measured INS spectrum at 7 K with $E_i$ = 28.2 meV for $\alpha$-$Cu_2P_2O_7$. (b) Simulated (using the SPINW program) spin wave spectrum with the model having $J_1$ = 2.8 meV, $J_2$ = 7.73 meV, $J_3$ = 3.61 meV, $J_4$ = 3.23 meV, $J_5$ = 0.03 meV, and $D$ = -0.07 meV (Table V). The calculated spin-wave spectrum is powder averaged, convoluted with the energy-transfer-dependent instrumental resolution, and corrected for the $Cu^{2+}$ magnetic form factor. (c) and (d) Comparison of experimental and calculated plots for constant-energy (integration over 2–4 meV) and constant-$|Q|$ (integration over $|Q|$ = 0–3 Å$^{-1}$) cuts. The experimentally observed data are shown by symbols, and the calculated curves are shown by red solid lines. Low energy (e) experimental INS spectrum ($E_i$ = 11.3 meV) and (f) calculated spectrum illustrating the two distinct energy gaps at $|Q|$ = 0.78 and 1.62 Å$^{-1}$. (g) and (h) Comparison of experimental and calculated plots (constant-$|Q|$ cuts) for $|Q|$ = 0.78 and 1.62 Å$^{-1}$ [integration over 0.75-0.85 and 1.4-1.8 Å$^{-1}$, respectively).

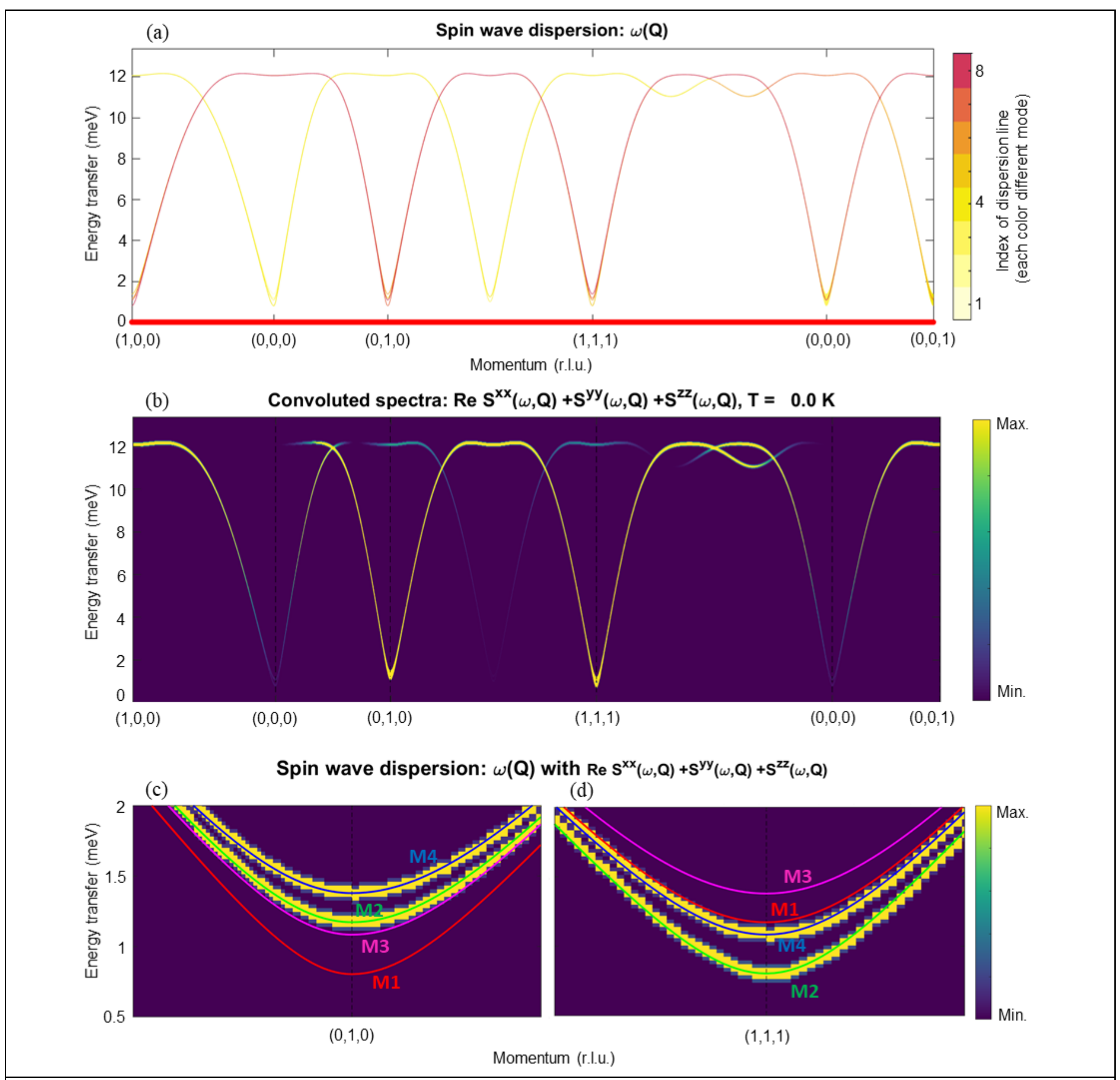

FIG. 9. (a) The simulated dispersion spectra (by the SPIN-W program) along the different crystallographic directions for the determined Hamiltonian (Table V) with parameters $J_1 = 2.8$ meV, $J_2 = 7.73$ meV, $J_3 = 3.61$ meV, $J_4 = 3.23$ meV, $J_5 = 0.03$ meV, and $D = -0.07$ meV. (b) The intensity variation of the dispersion spectra is shown by the color map. (c) and (d) The zoomed dispersion spectra around the AFM zone centres (0,1,0) and (1,1,1), respectively, highlighting four dispersion modes M1-M4.

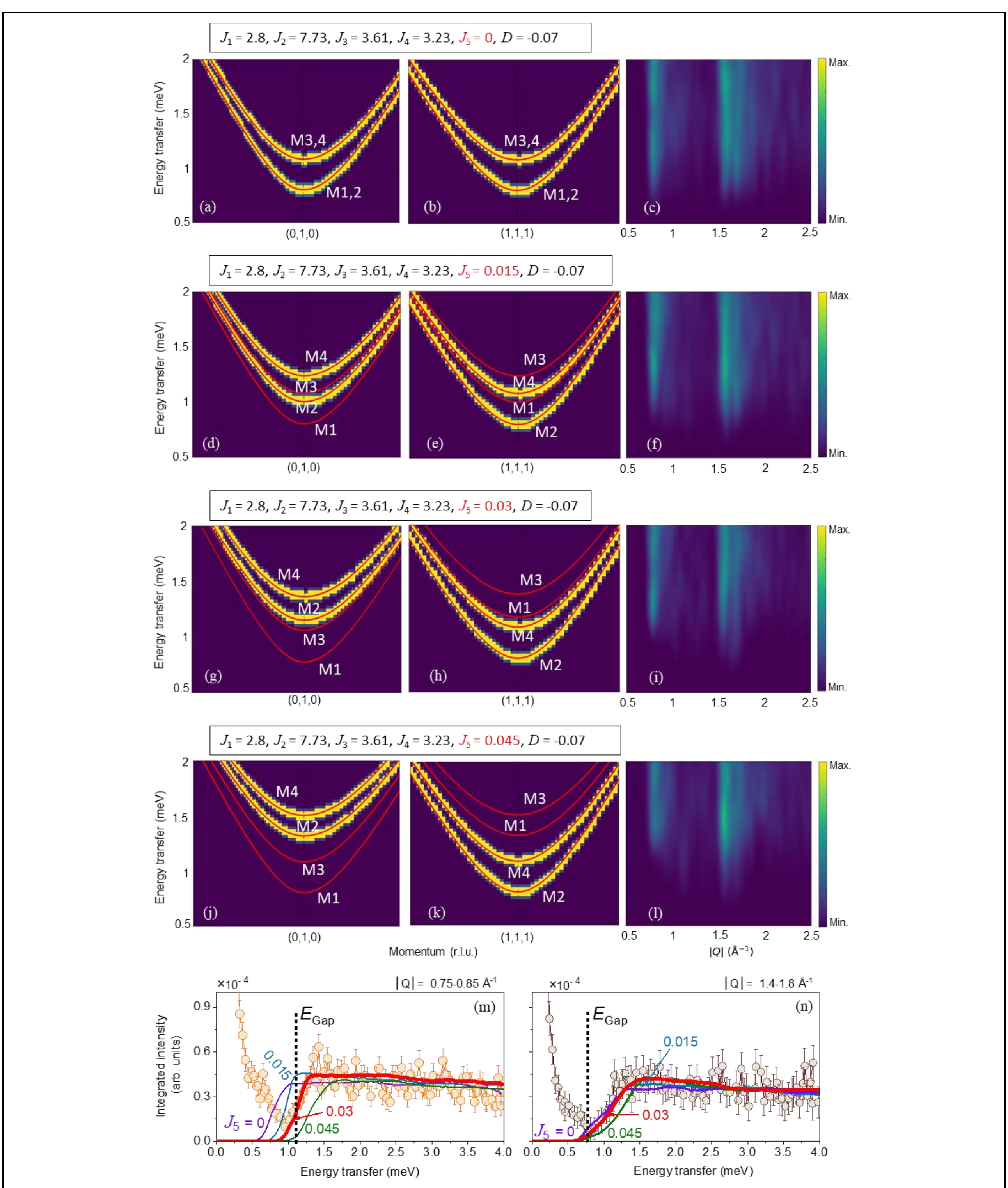

FIG. 10. Simulated magnetic excitation spectra (by the SPIN-W program) for different values of the inter-planar exchange interaction $J_5$ (0, 0.015, 0.03, and 0.045 meV) for (a-b, d-e, g-h and j-k) single crystal and for (c, f, i and l) powder samples, respectively. (m and n) The comparison of the simulated constant |$Q$| cuts with the same from experimental spectra for the AFM zone centres (0,1,0) [|$Q$| = 0.78 Å$^{-1}$] and (1,1,1) [|$Q$| = 1.62 Å$^{-1}$]. All the spectra are simulated with the fixed values of all other parameters as given in Table V and mentioned at the top of each panel.

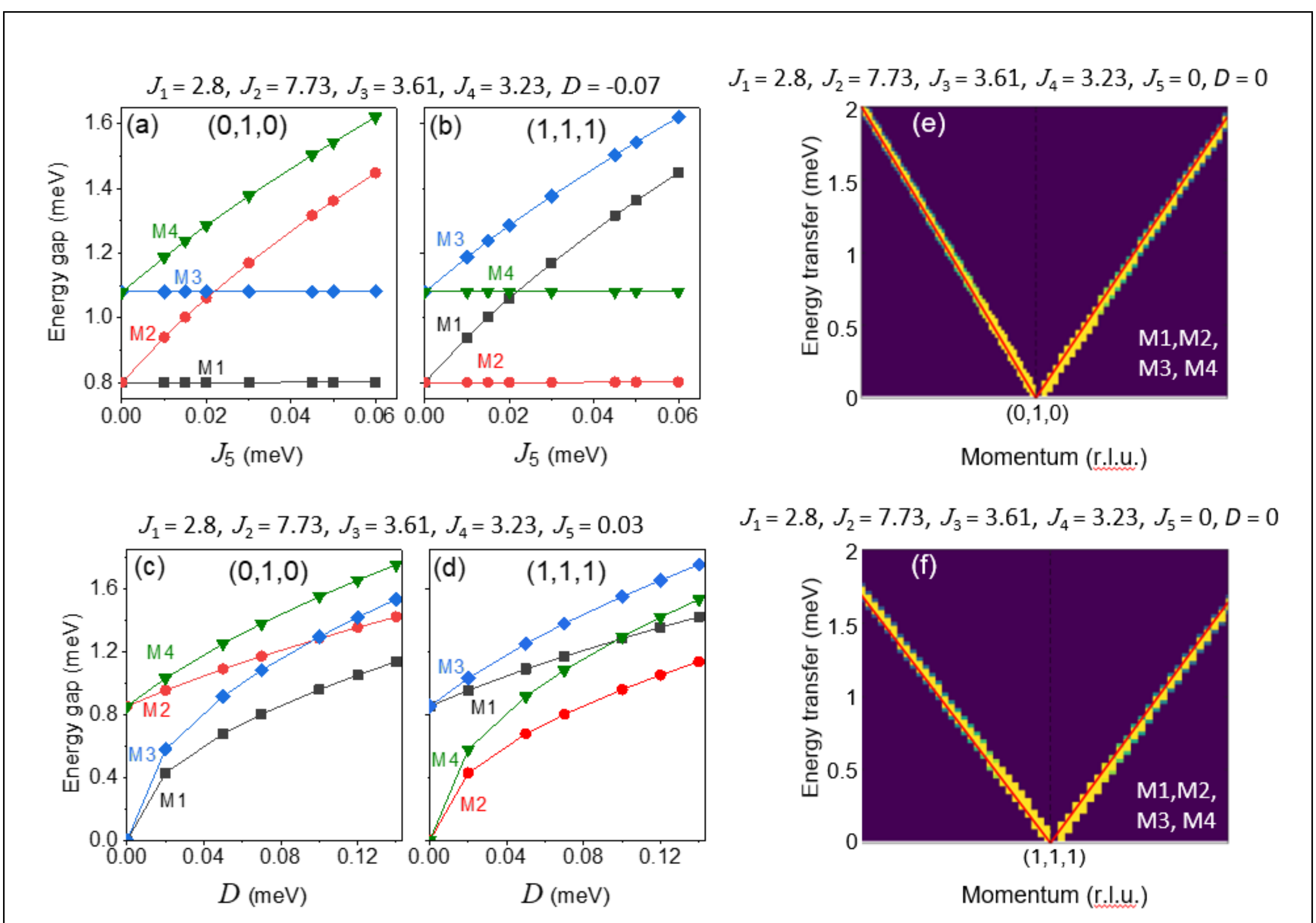

FIG. 11. The variations of the energy gap values as a function of (a, b) inter-planar exchange interaction $J_5$ and (c, d) single ion anisotropy parameter $D$ for two AFM zone centres (0,1,0) [$|Q|$ = 0.78 Å$^{-1}$] and (1,1,1) [$|Q|$ = 1.62 Å$^{-1}$], respectively. (e) and (f) Simulated magnetic excitation spectra (by the SPIN-W program) for $J_5$ = 0 and $D$ = 0 around the AFM zone centres (0,1,0) and (1,1,1), respectively.

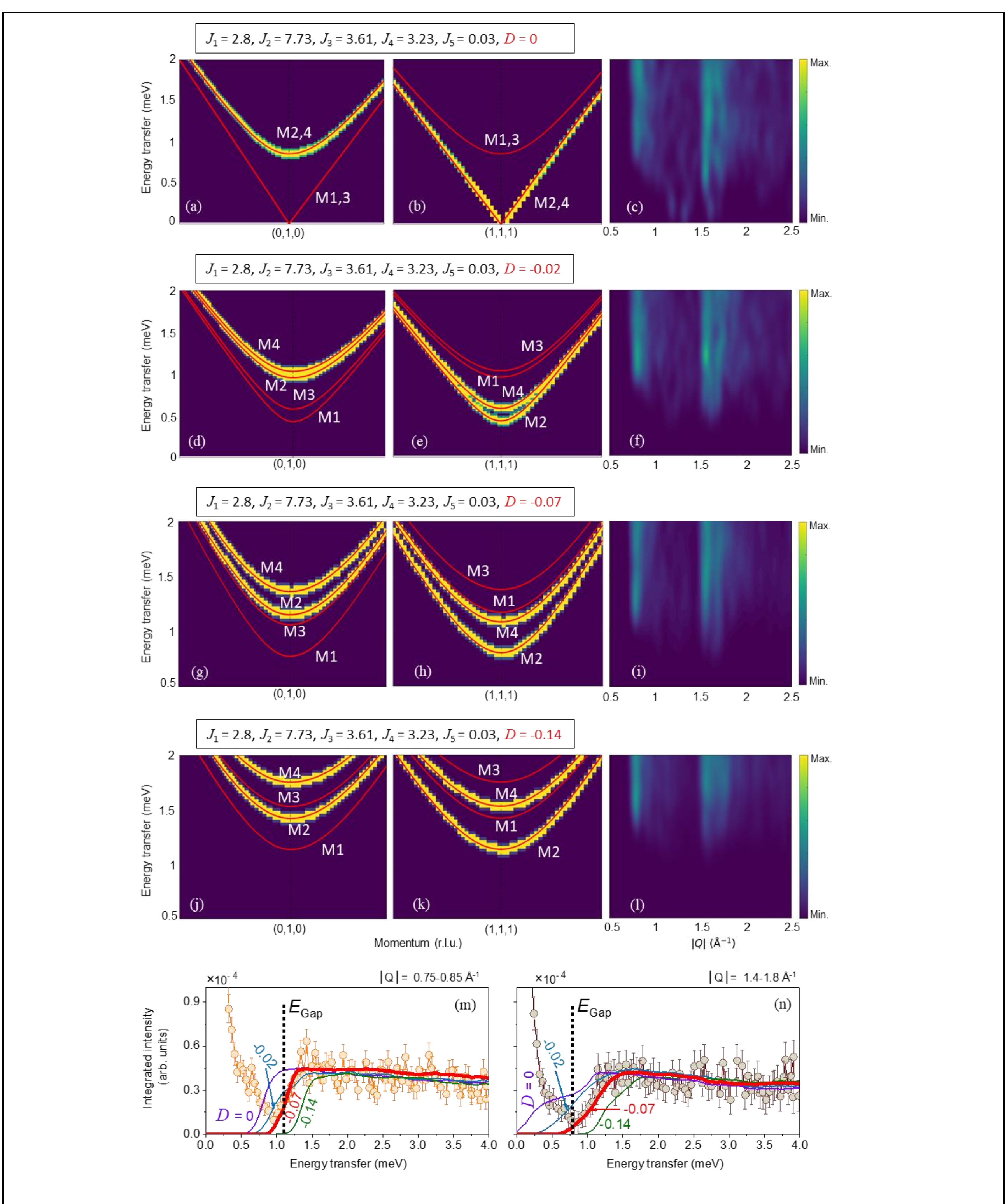

FIG. 12. Simulated magnetic excitation spectra (by the SPIN-W program) for different values of the single ion anisotropy parameter $D$ (0, -0.02, -0.07, and -0.14 meV) for (a-b, d-e, g-h and j-k) single crystal and for (c, f, i and l) powder samples, respectively. (m and n) The comparison of the simulated constant $|Q|$ cuts with the same from experimental spectra for the AFM zone centres (0,1,0) [$|Q|$ = 0.78 Å$^{-1}$] and (1,1,1) [$|Q|$ = 1.62 Å$^{-1}$]. All the spectra are simulated with the fixed values of all other parameters as given in Table V and mentioned at the top of each panel.

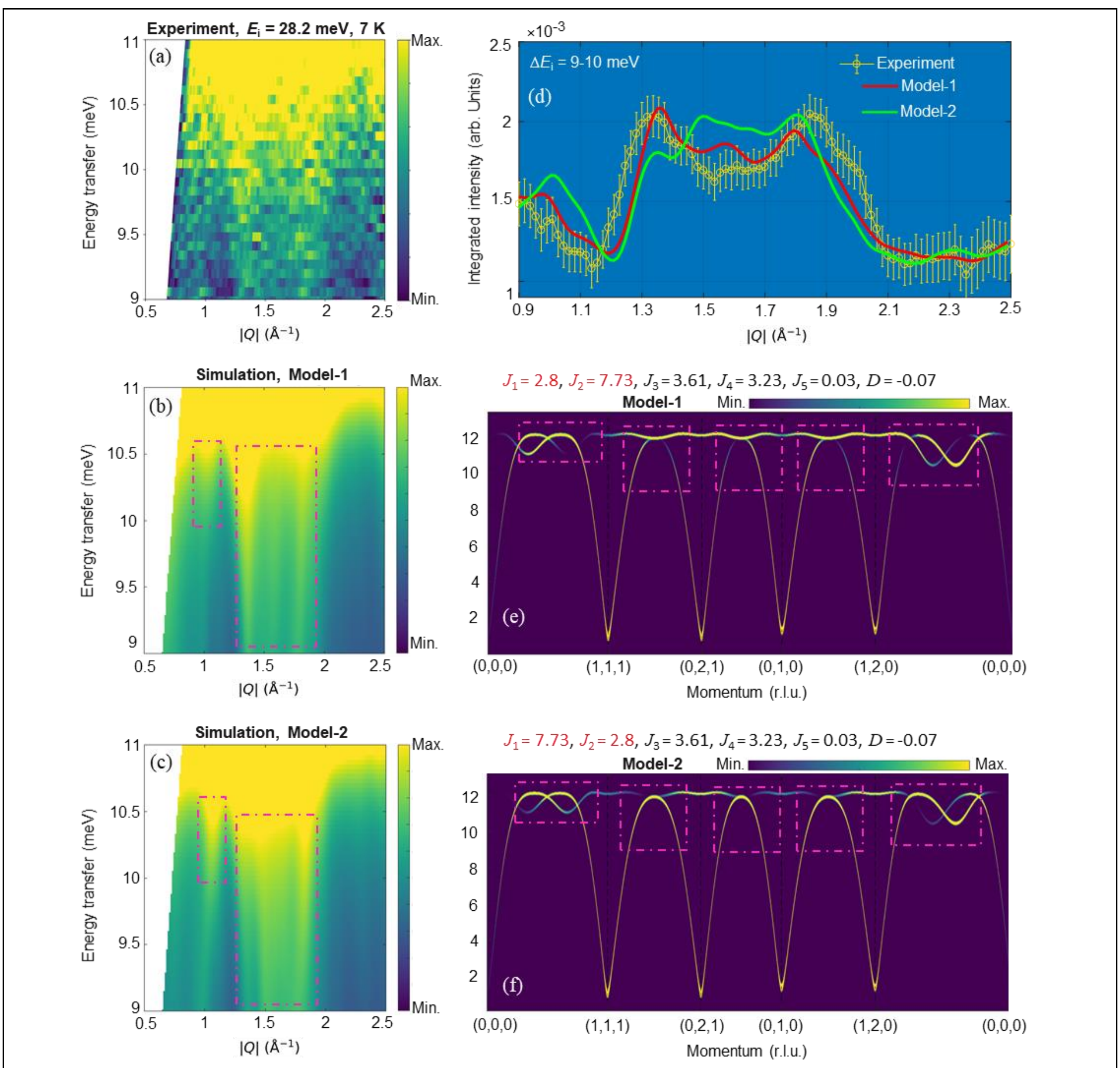

FIG. 13. (a) Experimentally measured (at 7 K with $E_i$ = 28.2 meV) INS intensity for $\alpha$-$Cu_2P_2O_7$ zoomed over 9-11 meV and 0.5-2.5 Å$^{-1}$. (b, c) Simulated (using the SPINW program) spin-wave excitation spectra for two spin-dimer models; Model-1 (structural dimer) and Model-2 (magnetic dimer). (d) Comparison of the experimental intensity vs momentum transfer ($|Q|$) data (symbols) with the calculated curves considering the Model-1 (red curve) and Model-2 (green curve). (e) and (f) The simulated intensity variation of the dispersion patterns along the different crystallographic directions for Model-1 and Model-2, respectively. The dotted boxes highlight the differences in the simulated spectra for two models.

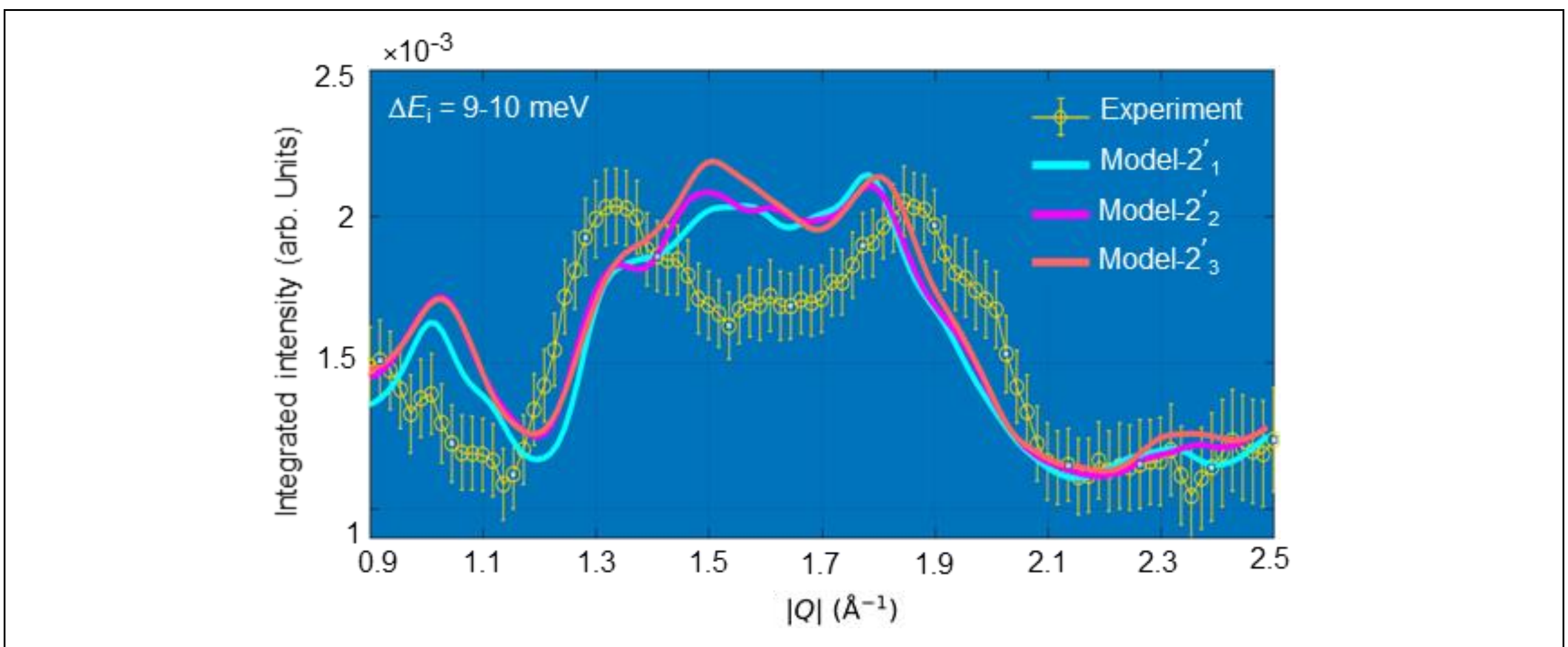

FIG. 14. Comparison of the experimental intensity vs momentum transfer ($|Q|$) data (symbols), obtained by integration over the energy range 9-10 meV of the measured spectra at 7 K with $E_i$ = 28.2 meV, with the calculated curves considering the Model-$2_1'$ , Model-$2_2'$ and Model-$2_3'$.